\documentclass[10pt,letterpaper]{article}
\usepackage[top=0.85in,left=2.75in,footskip=0.75in]{geometry}

\usepackage{amsmath,amssymb}

\usepackage{changepage}

\usepackage[utf8x]{inputenc}

\usepackage{textcomp,marvosym}

\usepackage{cite}

\usepackage{nameref,hyperref}

\usepackage[right]{lineno}

\usepackage{microtype}
\DisableLigatures[f]{encoding = *, family = * }

\usepackage[table]{xcolor}

\usepackage{array}

\newcolumntype{+}{!{\vrule width 2pt}}

\newlength\savedwidth

\raggedright
\usepackage[aboveskip=1pt,labelfont=bf,labelsep=period,justification=raggedright,singlelinecheck=off]{caption}

\makeatletter
\renewcommand{\@biblabel}[1]{\quad#1.}
\makeatother

\usepackage{lastpage,fancyhdr,graphicx}
\usepackage{epstopdf}
\fancyheadoffset[L]{2.25in}
\fancyfootoffset[L]{2.25in}
\begin{document}
\vspace*{0.2in}

\begin{flushleft}
{\Large 
\textbf\newline{MSPM: A modularized and scalable multi-agent reinforcement learning-based system for financial portfolio management} 
}
\newline
\\
Zhenhan Huang\textsuperscript{1*},
Fumihide Tanaka\textsuperscript{2}
\\
\bigskip
\textbf{1} Graduate School of Science and Technology, University of Tsukuba, Tsukuba, Ibaraki, Japan
\\
\textbf{2} Faculty of Engineering, Information and Systems, University of Tsukuba, Ibaraki, Japan
\\
\bigskip

* Corresponding author \\
E-mail: huang@ftl.iit.tsukuba.ac.jp (ZH)

\end{flushleft}

\section*{Abstract}
Financial portfolio management (PM) is one of the most applicable problems in reinforcement learning (RL) owing to its sequential decision-making nature. However, existing RL-based approaches rarely focus on scalability or reusability to adapt to the ever-changing markets. These approaches are rigid and unscalable to accommodate the varying number of assets of portfolios and increasing need for heterogeneous data input. Also, RL agents in the existing systems are ad-hoc trained and hardly reusable for different portfolios. To confront the above problems, a modular design is desired for the systems to be compatible with reusable asset-dedicated agents. In this paper, we propose a multi-agent RL-based system for PM (MSPM). MSPM involves two types of asynchronously-updated modules: Evolving Agent Module (EAM) and Strategic Agent Module (SAM). An EAM is an information-generating module with a Deep Q-network (DQN) agent, and it receives heterogeneous data and generates signal-comprised information for a particular asset. An SAM is a decision-making module with a Proximal Policy Optimization (PPO) agent for portfolio optimization, and it connects to multiple EAMs to reallocate the corresponding assets in a financial portfolio. Once been trained, EAMs can be connected to any SAM at will, like assembling LEGO blocks. With its modularized architecture, the multi-step condensation of volatile market information, and the reusable design of EAM, MSPM simultaneously addresses the two challenges in RL-based PM: scalability and reusability. Experiments on 8-year U.S. stock market data prove the effectiveness of MSPM in profit accumulation by its outperformance over five different baselines in terms of accumulated rate of return (ARR), daily rate of return (DRR), and Sortino ratio (SR). MSPM improves ARR by at least 186.5\% compared to constant rebalanced portfolio (CRP), a widely-used PM strategy. To validate the indispensability of EAM, we back-test and compare MSPMs on four different portfolios. EAM-enabled MSPMs improve ARR by at least 1341.8\% compared to EAM-disabled MSPMs.

\section*{Introduction}
Portfolio management (PM) is a continuous process of reallocating capital into multiple assets \cite{10.2307/2975974}, and it aims to maximize accumulated profits with an option to minimize the overall risks of the portfolio. To perform such a practice, portfolio managers who focus on stock markets conventionally read financial statements and balance sheets, follow the news from media and announcements from financial institutions and analyze stock price trends. By the resemblant nature of the problem, researchers expectedly wish to incorporate deep reinforcement learning (DRL) methods in PM. As one of the attempts, the authors of \cite{jiang2017deep} propose a PM framework for cryptocurrencies using Deep Deterministic Policy Gradient (DDPG) \cite{10.5555/3044805.3044850,lillicrap2019continuous}. \cite{liang2018adversarial} proposes a method called Adversarial Training for portfolio optimization with the implementation of three different RL methods: DDPG, Proximal Policy Optimization (PPO)\cite{schulman2017proximal} and Policy Gradient (PG). Akin to receiving information from various sources as portfolio managers generally do, existing approaches incorporate heterogeneous data \cite{Ye_Pei_Wang_Chen_Zhu_Xiao_Li_2020}. Recently, multi-agent reinforcement learning (MARL) approaches are also proposed by researchers\cite{10.1007/978-3-642-12433-4_7,Sycara-1995-14017,Lee_2020}. In\cite{Lee_2020}, the authors propose MAPS, a system involving a group of Deep Q-network \cite{mnih2013atari} (DQN)-based agents corresponding to individual investors, to make investment decisions and create a diversified portfolio. MAPS can be recognized as a reinforcement-learning implementation of ensemble learning \cite{ensemble} by its very nature. In addition,\cite{Liu_Liu_Zhao_Pan_Liu_2020} proposes iRDPG to generate adaptive quantitative trading strategies by using DRL and imitation learning. However, while inspiring, the existing approaches seldom focus on scalability and reusability to accommodate the ever-changing markets. RL agents in the existing multi-agent-based systems are ad-hoc trained and rarely reusable for different portfolios. Also, the existing systems are barely scalable to answer the need for scaled number of assets in portfolios and increasing heterogeneous data input. For example, in SARL \cite{Ye_Pei_Wang_Chen_Zhu_Xiao_Li_2020}, the encoder's intake is either financial news data for embedding or stock prices for trading signals generation, but can not be both of them, and this issue prevents the encoder from efficiently producing holistic information and eventually limits the RL-based agents' learning. Furthermore, the existing systems lack a modular design to be compatible with different RL agents for different assets.
In this paper, we propose MSPM, a novel multi-agent reinforcement learning-based system, with a modularized and scalable architecture for PM. In MSPM, assets are vital and organic building blocks. This vitalness is reflected in that each asset has its dedicated module: Evolving Agent Module (EAM). An EAM takes heterogeneous data and utilizes a DQN-based agent to produce signal-comprised information. After we set up and trained the EAMs corresponding to the assets in a portfolio, we connected them to a decision-making module: Strategic Agent Module (SAM). An SAM represents a portfolio and uses the profound information from the connected EAMs for asset reallocation. EAM and SAM are asynchronously updated, and EAMs' reusability allows themselves to be combined and connected to multiple SAMs discretionarily. With the power of parallel computing, we can perform capital reallocation for various portfolios at scale, simultaneously.\\
To evaluate MSPM's performance, we back-test and compare MSPM to five different baselines on two different portfolios. MSPM outperforms all the baselines in terms of accumulated rate of return, daily rate of return, and Sortino ratio. For instance, MSPM improves accumulated rate of return by 49.3\% and 426.6\% compared to the state-of-the-art RL-based method: Adversarial PG\cite{liang2018adversarial} on the two portfolios. We also inspect the position-holding of five different EAMs to exemplify the high quality and reliability of the signals generated by EAM. Specifically, the average winning rate of the EAMs in the two portfolios achieves 80\%. Furthermore, we validate the necessity of EAM by back-testing and comparing the EAM-enabled and disabled MSPMs on four different portfolios. EAM-enabled MSPMs improve accumulated rate of return by at least 1341.8\% compared to the EAM-disabled MSPMs. Our contribution can be listed as follows:
\begin{itemize}
    \item To the best of our knowledge, MSPM is the first approach that formalizes a modularized and scalable multi-agent reinforcement learning system using signal-comprised information for financial portfolio management.
    \item MSPM with its modularized and reusable design addresses the issue of ad-hoc, fixed, and inefficient model training in the existing RL-based methods.
    \item By experiment and comparison, we confirm that our MSPM system outperforms five different baselines under extreme market conditions of U.S. stock markets during the global pandemic, from January to December 2020.
    \item EAM-enabled MSPM systems improve accumulated rate of return of two different portfolios by 49.3\% and 426.6\% compared to Adversarial PG\cite{liang2018adversarial}, a state-of-the-art RL-based method, and by 186.5\% and 369.8\% compared to Constant Rebalanced Portfolio (CRP)\cite{10.1093/rfs/hhm075}, a conventional PM strategy. In addition, the average winning rate of the EAMs in the two portfolios achieves 80\%.
    \item Furthermore, we validate the indispensability of Evolving Agent Module (EAM) by back-testing MSPM on four different investment portfolios. Among the portfolios, EAM-enabled MSPMs improve accumulated rate of return by at least 1341.8\% compared to the EAM-disabled MSPMs.
    
\end{itemize}

\section*{Related work}
In the early years, researchers and professionals believe that certain behaviors of price and volume will repeat periodically and consistently. Based on this recognition, the technical indicators (TI) are invented by using historical price and volume data to predict the movement of asset prices\cite{Fama1970EFFICIENTCM}. TIs are mostly formulas or particular patterns, and the trading strategies that utilize TIs are referred to as technical analysis (TA)\cite{TechnicalAnalysis}. However, as pre-defined formulas and patterns cannot cover all market movements, it is getting harder and harder for TA to adapt to the fast-changing market.
With the increase in computing power and available data, researchers have started to use deep learning (DL) to predict stock price movements. DL uses high-dimensional data to train complex and non-linear neural network models as trading strategies. Fortunately, DL's adaptability to the market is promisingly improved compared to TA.
Recently, deep reinforcement learning (DRL) has emerged rapidly as the combination of DL and reinforcement learning (RL). By utilizing neural networks (NN), a DRL-based agent is particularly good at extracting useful information from high-dimensional data and taking sequential actions based on rewarding. DRL methods have led to many breakthroughs in multiple fields. For instance,\cite{mnih2013atari} successfully utilizes Deep Q-learning agents to learn directly from high-dimensional raw pixel input to play video games. Due to the sequential decision-making nature of financial investment, researchers naturally attempt to solve stock trading problems using DRL methods. \cite{jiang2017deep} designed a cryptocurrencies portfolio management (PM) framework using Deep Deterministic Policy Gradient (DDPG)\cite{10.5555/3044805.3044850,lillicrap2019continuous} which is a model-free DRL algorithm. \cite{liang2018adversarial} proposes the Adversarial Training method to improve training efficiency using three different RL methods: DDPG, Proximal Policy Optimization (PPO)\cite{schulman2017proximal} and Policy Gradient (PG). Although these approaches have presented potential performance, the data input of these approaches is still traditional historical data, namely opening-high-low-closing prices (OHLC) and trading volumes. Unlike preceding research, \cite{Ye_Pei_Wang_Chen_Zhu_Xiao_Li_2020} proposes SARL, an RL framework that can incorporate heterogeneous data to generate PM strategies. Moreover, to address the challenge of balancing between exploration and exploitation, \cite{Liu_Liu_Zhao_Pan_Liu_2020} proposes iRDPG for developing trading strategies by DRL and imitation learning. Multi-agent systems have also been proposed. In\cite{Lee_2020}, the authors propose MAPS, a cooperative system containing multiple agents, to create diversified portfolios and to adapt to the continuously changing market conditions.
However, while the existing approaches tackle PM problems with promising methods and techniques, these systems, with the strategies generated, are mostly fixed and ad-hoc. The existing systems or frameworks lack a modular design to be compatible with different trained RL agents. The RL agents trained for one portfolio can hardly be reused for different portfolios. These systems also lack scalability to accommodate the increasing number of assets and profundity of market information. In this paper, we propose MSPM for solving the problems.

\section*{Data}
\subsection*{Data acquisition}
The historical price data used in this paper are QuoteMedia’s End of Day US Stock Prices (EOD) \cite{QuoteMedia} from Jan 2013 to Dec 2020 obtained using Nasdaq Data Link's API, which can be accessed by subscribing at: \url{https://data.nasdaq.com/data/EOD-end-of-day-us-stock-prices}. We also use web news sentiment data (FinSentS)\cite{NS1} from Nasdaq Data Link provided by InfoTrie, which can be accessed by subscribing at: \url{https://data.nasdaq.com/databases/NS1/data}.

\subsection*{Feature selection and data curation}
We select the adjusted- close, open, high, and low prices and volumes features from QuoteMedia’s EOD data as the historical price data. We also select the sentiment and news\_buzz from InfoTrie's FinSentS Web News Sentiment. Each feature in EOD data is normalized by dividing the first (day-one) value of that feature, and there is no missing value in any of these features. For FinSentS data, we use original values of the sentiment feature in FinSentS data, and we fill the missing values (accounting for 9.51\% of the total data) prior year 2013 with a neutral sentiment: zero (0). Since the FinSentS data are not as straightforward as EOD data, we put the description of the selected features of FinSentS data in Table~\ref{tab:sent_data}.

\begin{table}[!ht]
\centering
\caption{Feature selection of FinSentS data}
\begin{tabular}{@{}llll@{}}
\hline\noalign{\smallskip}
Feature & Description \\
\noalign{\smallskip}\hline\noalign{\smallskip}
sentiment & A measure of bullishness and bearishness of equity prices calculated as\\
& a statistical index of the news corpus. Sentiment scores are defined\\
& on a scale of -5 to 5 indicating from the most bearish to the most bullish.\\
\noalign{\smallskip}\hline\noalign{\smallskip}
news\_buzz & Normalized value of change in standard deviations of periodic number\\
& of news items (news volume) used for generating sentiment. Buzz scores \\
& reflect a sharp change in news volume thus serving as a risk alert\\
& indicator. Defined on a scale of 1-10 high buzz score reflects higher\\
& volatility.\\
\noalign{\smallskip}\hline
\end{tabular}
\label{tab:sent_data}
\end{table}

\section*{Methodology}
\begin{figure}[!h]
        \centering
         \includegraphics[width=\linewidth]{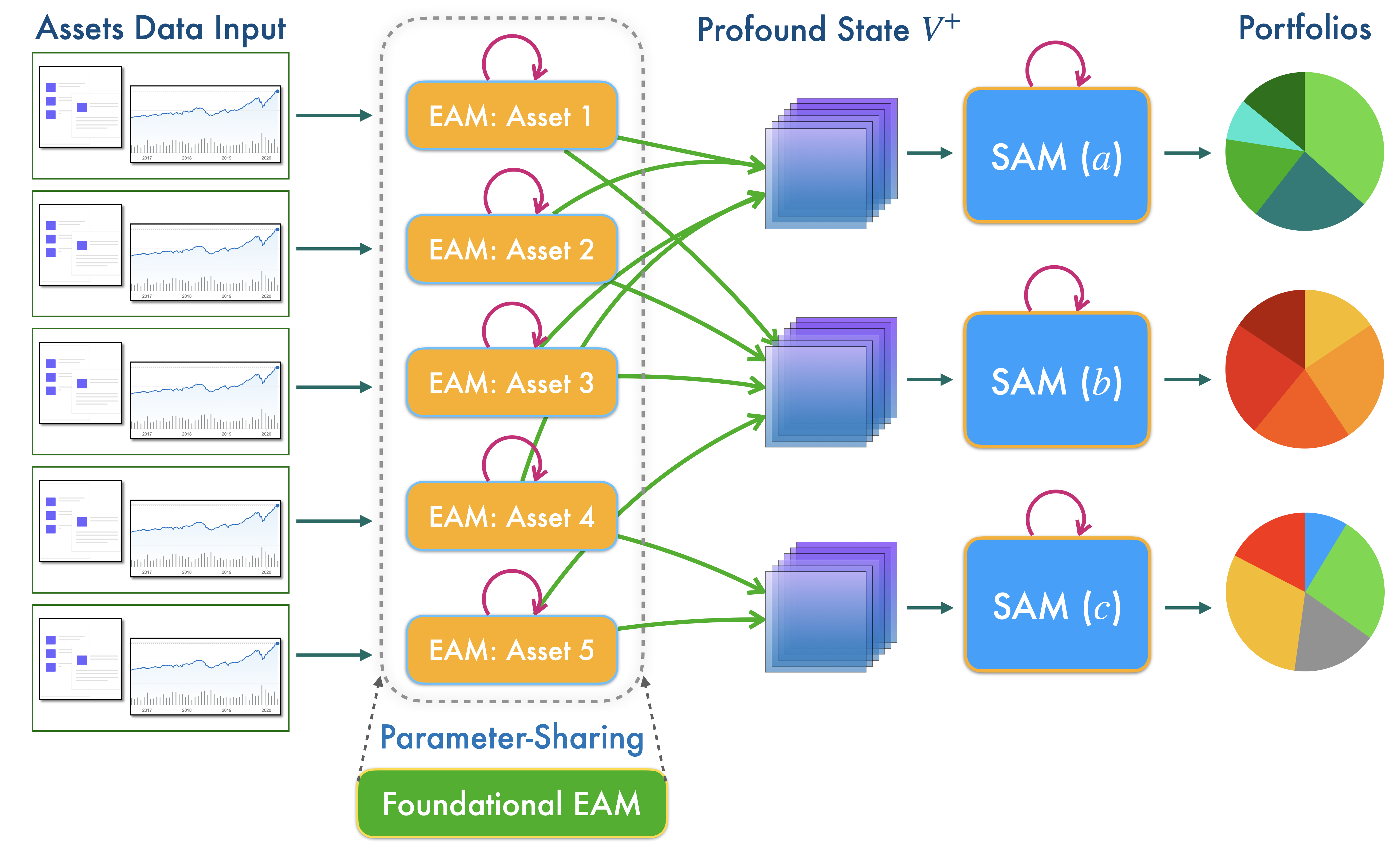}
               \caption{{\bf Overview of the surjection relationship between Evolving Agent Modules (EAMs) and Strategic Agent Modules (SAMs)} Each EAM is responsible for a single asset and employs a DQN agent, and it utilizes heterogeneous data to produce signal-comprised information. Each SAM is a module for a portfolio that employs a PPO agent to reallocate the assets with stacked signal-comprised 3-D tensor \textbf{\textit{profound state}} $V^+$ from EAMs connected. Moreover, trained EAMs are reusable for different portfolios and therefore can be combined and connected to any SAMs at will. By parallel computing, capital reallocation may be performed for various portfolios at scale simultaneously.}
          \label{fig:fig1}
    \end{figure}

\begin{figure}[!h]
        \centering
         \includegraphics[width=\linewidth]{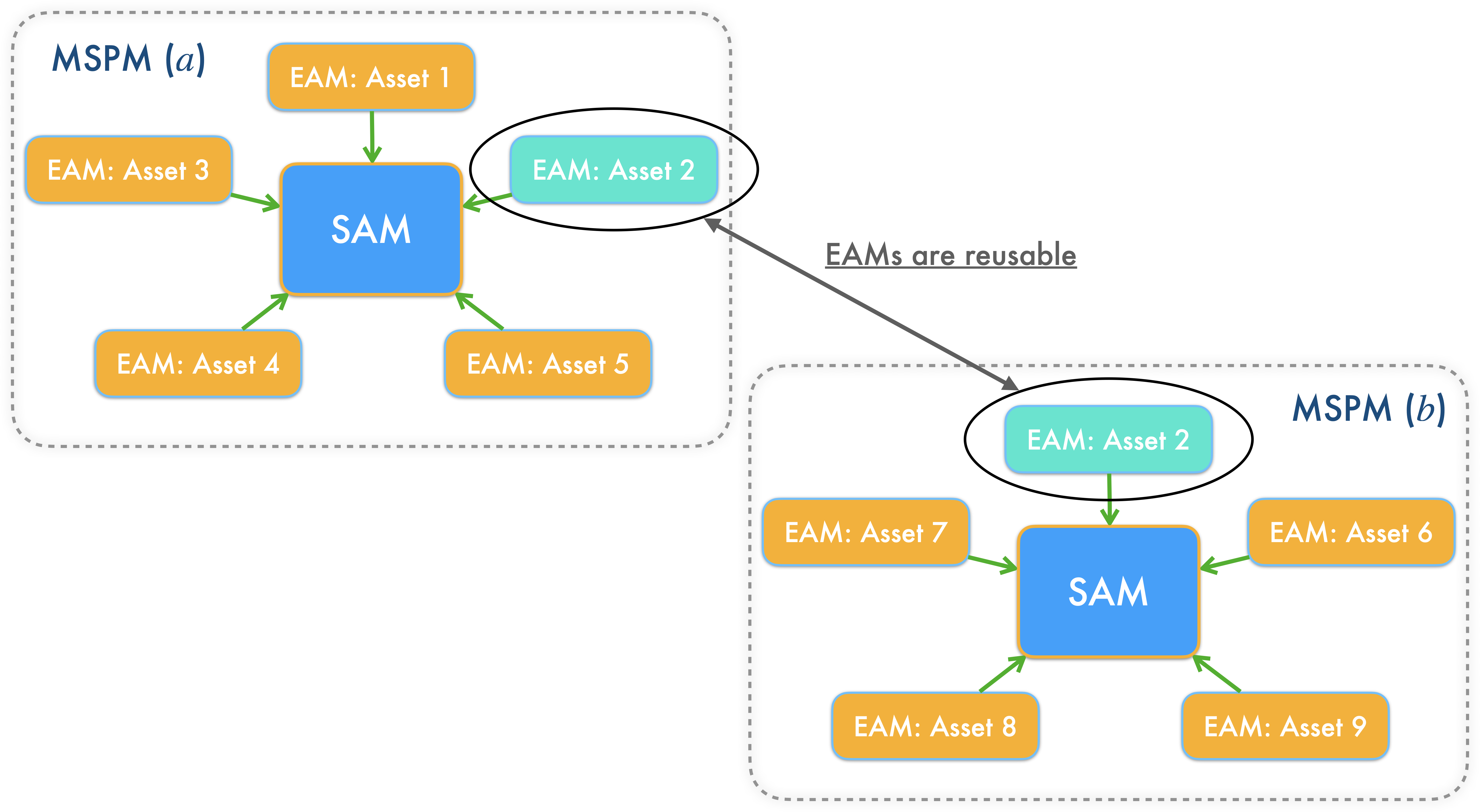}
               \caption{{\bf A more intuitive illustration of MSPM's architecture} EAMs are reusable for different portfolios. EAMs can be combined and connected to any SAMs at will, like assembling LEGO blocks.}
          \label{fig:relationship_2}
    \end{figure}

\begin{figure}[!h]
        \centering
           \includegraphics[width=0.75\linewidth]{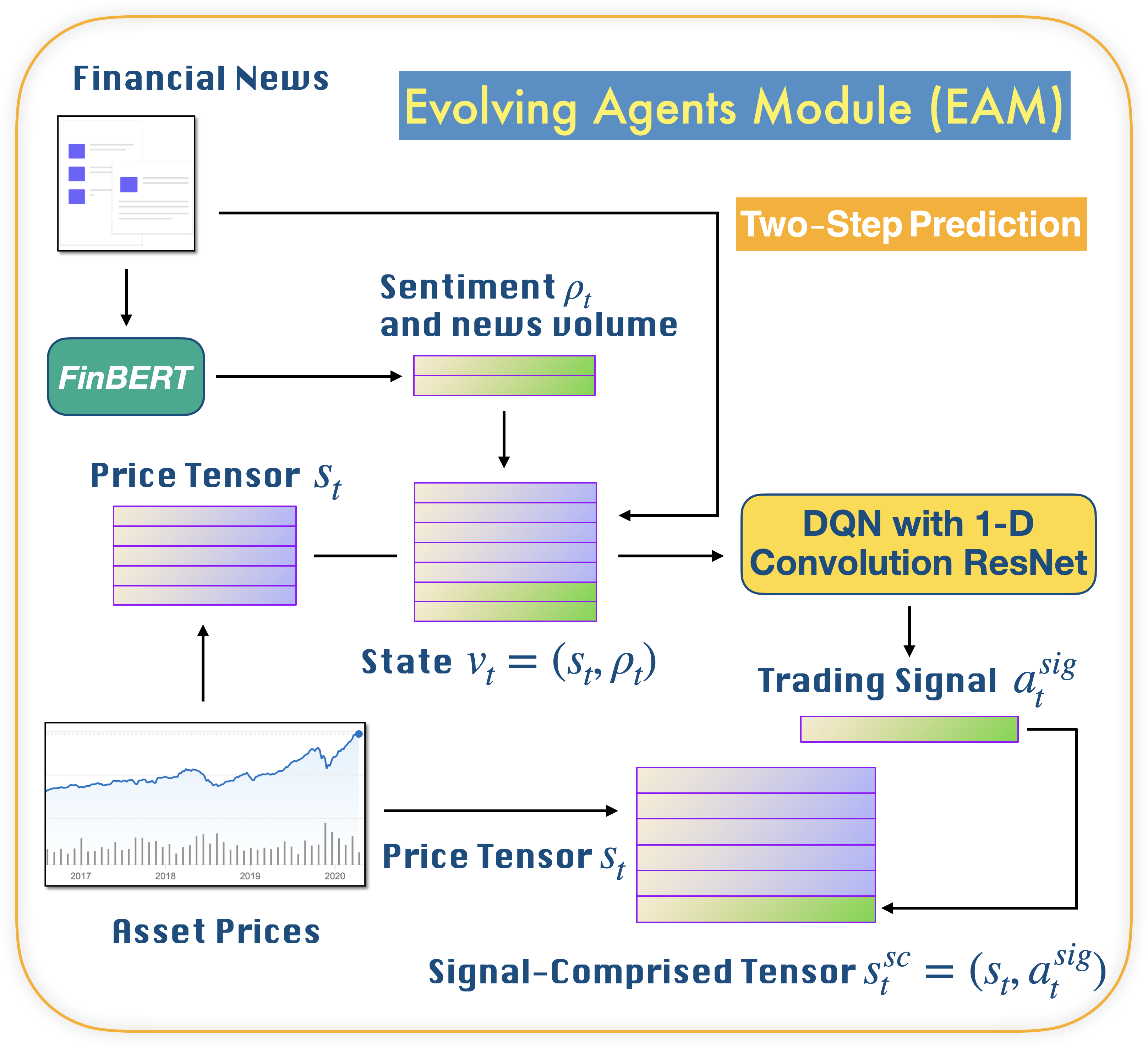}
               \caption{{\bf Abstract of EAM's architecture.} An EAM is a module for a designated asset. Each EAM takes two types of heterogeneous data: 1. designated asset’s historical prices and 2. asset-related financial news. At the center of an EAM is an extended DQN agent using a 1-D convolution ResNet for sequential decision making. Instead of training every EAM from scratch, we train EAMs by transfer learning using a foundational EAM. At every time step $t$, the DQN agent in EAM observes state $v_t$ of historical prices $s_t$ and news sentiments $\rho_t$ of the designated asset, and acts to trade with an action $a^{sig}_t$ of either buying, selling, or skipping, and eventually generates a 2-D signal-comprised tensor $s^{sc}_t$ using new prices $s_t$ and signals $a^{sig}_t$.}
          \label{fig:eam}
    \end{figure}

\begin{figure}[!h]
        \centering
           \includegraphics[width=0.55\linewidth]{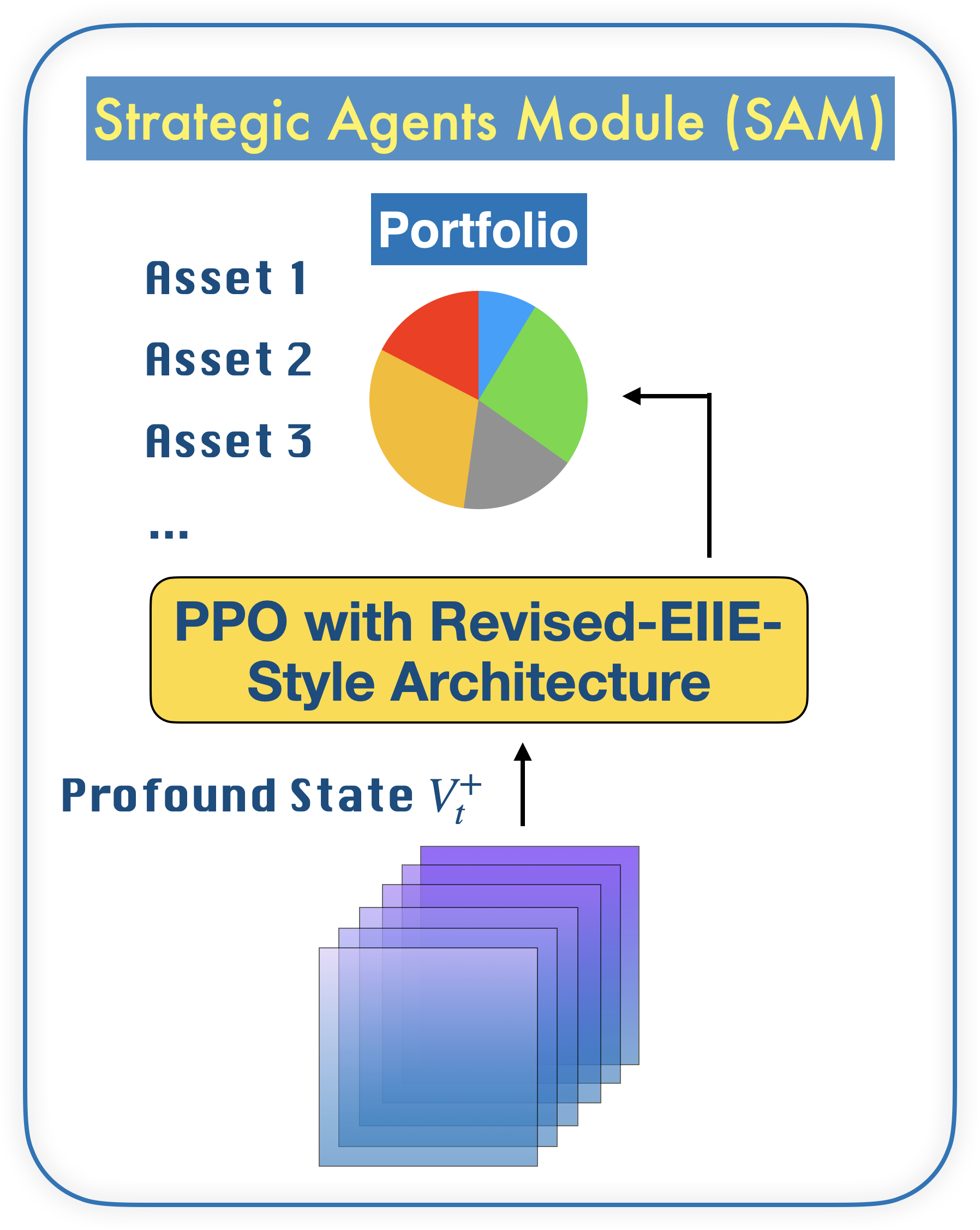}
               \caption{{\bf Abstract of SAM's architecture.} An SAM is a module for an investment portfolio. The input of SAM, profound state $V^+_t \in \mathbb{R}^{f \times m^* \times n}$, is a 3-D tensor, where $f$ is the number of features, $m^*=m+1$ is the number of assets $m$ in the portfolio plus cash and $n$ is the fixed rolling-window length. Each SAM takes the profound state $V^+_t$ which is stacked and transformed from 2-D tensors from connected EAMs, and further generates the reallocation weights for the assets in the portfolio.}
          \label{fig:sam}
    \end{figure}

Our MSPM system consists of two types of modules: EAM and SAM. The relationship between EAMs and SAMs is illustrated in Fig~\ref{fig:fig1}. Fig~\ref{fig:relationship_2} illustrates a even more intuitive overview of MSPM's architecture. To accommodate MSPM in the sequential decision-making problems financial portfolio management, we configured the specific settings for EAM and SAM. An EAM contains a DQN \textbf{agent} and \textbf{acts} to generate signal-comprised information (historical prices with buy/closing/skip labels) for a designated asset. To train the agent in EAM, we constructed a sequential decision-making problem with designated asset’s historical prices and financial news as the \textbf{state} that the agent observes at each time step. An DQN agent acts to buy or close a position, or simply to skip at every time step based on the latest prices and financial news data input, in order to maximize its total reward. The actions (signals) then will be matched and stacked back to the corresponding price data to formalize the signal-comprised information. EAM's architecture is illustrated in Fig~\ref{fig:eam}.
On the other hand, an SAM manages an investment portfolio and contains a PPO \textbf{agent} that reallocates the assets in that portfolio. SAMs are connected to multiple EAMs as an investment portfolio often has more than one asset. In the decision-making process of SAM, the state that the PPO agent \textbf{observes} at each time step is the combination of the signal-comprised information which the connected EAMs generate. Further, the PPO agent \textbf{acts} to generate the reallocation weights for the assets in the portfolio, which total up to 1.0. Fig~\ref{fig:sam} provides an overview of the SAM's architecture.
For both EAM and SAM, the composition of the assets’ historical prices and financial news or news sentiments is the \textbf{environment} their agents interact with. Each EAM is reusable. Once an EAM is set up and trained, it can be effortlessly connected to any SAM. An SAM connects to at least one EAM. EAMs are retrained periodically using the latest information from the market, media, financial institutions, etc., and we implemented the former two data sources in this study.
In the following sections, we explain the technical details of EAM and SAM.

\subsection*{Evolving Agent Module (EAM)}

\subsubsection*{State}
At any given periodic (daily) time-step $t$, the agent in EAM observes state $v_t$, which consists of the designated asset’s recent $n$-day historical prices $s_t$ and sentiment scores $\rho_{t}$. Specifically,
\begin{equation}
v_t = (s_t, \rho_{t}),
\end{equation}
where $s$ includes the designated asset’s $n$-day close, open, high and low prices and volumes. $\rho$ includes the predicted and averaged news sentiments, using a pre-trained FinBERT classifier \cite{araci2019finbert,devlin-etal-2019-bert} for asset-related financial news, which ranges continuously from -5.0 to 5.0, indicating bearishness (-5.0) or bullishness (5.0).
Furthermore, $\rho$ also includes \textit{news\_buzz}. This attribute is an attempt to alleviate the unbalanced-news issue in the existing research \cite{Ye_Pei_Wang_Chen_Zhu_Xiao_Li_2020}.
Instead of restarting from the beginning after every episodic reset of the environment, the environment resets at a random time point of the data \cite{10.5555/3279266}.\\
Because the news sentiments from FinSentS data and the sentiments generated by FinBERT are similar, and due to the restriction of APIs and web scraping, we only utilize FinSentS data as the sentiments input for the experiments in this paper.

\subsubsection*{Deep Q-Network}
For an EAM, we train a Deep Q-network (DQN) agent and follow the sequential decision-making of Deep Q-learning \cite{mnih2013atari}. Deep Q-learning is a value-based method that derives a deterministic policy $\pi(\theta)$, which is a mapping: $S \rightarrow A$ from state space to discrete action space. We use a Residual Network with 1-D convolution \cite{8331585} to represent the state-value function $Q^{\theta}$ which the agent acts based on: 
\begin{equation}
Q^{\theta}(s_t,a_t) = \mathbb{E}_{  \pi_{\theta}}[\sum_{k=0}^{\infty}\gamma^{k}r_{t+k+1}\mid s_t=s,a_t=a]
\end{equation}
For information about model selection for EAM and hyperparameter tuning, see~\nameref{S1_Appendix}.

\textbf{DQN Extensions:}
We implement three extensions \cite{10.5555/3279266} of the original DQN, namely dueling architecture \cite{wang2016dueling}, Double DQN \cite{10.5555/3016100.3016191} and two-step Bellman unrolling.

\textbf{Transfer Learning:}
Instead of training every EAM from scratch, we initiate and train a \textbf{foundational EAM}, using historical prices of AAPL (Apple Inc.), and then train all other EAMs based on this pre-trained EAM. By doing so, the foundational EAM shares its parameters with other EAMs which obtains prior knowledge of the pattern of stock trends. This transfer learning approach may help to tackle the \textbf{data-shortage} issue of newly-listed stocks due to the limited historical prices and news data available for training purposes.

\subsubsection*{Action}
The DQN agent in EAM acts to trade the designated asset with an action of either buying, selling, or skipping, at every time step $t$. The choice of an action, $a_{t}$ = \{buying, closing, or skipping\}, is called an \textbf{\textit{asset trading signal}}. As indicated in the actions, there is no short (selling) position, and a new position will be opened only after an existing position has been closed.

\subsubsection*{Reward}
The reward, $r_t$, received by the DQN agent at each time step $t$ is:
\begin{equation}
r_t(s_t,\iota_t)
= \begin{cases}
 100 (\sum_{i=t_\iota}^{t}\frac{v^{(close)}_t}
 {v^{(close)}_{t-1}}-1-\beta) \text{, if } \iota_t\\
 0 \text{, if not } \iota_t
\end{cases}
\end{equation}
where $v^{(close)}_t$ is the close price of the given asset at time step $t$. $t_l$ is the time step when a long position is opened and commissions are deducted, $\beta$ stands for the commission of 0.0025 and $\iota_t$ is the indicator of an opening position (i.e., a position is still open).

\subsection*{Strategic agent module (SAM)}

\subsubsection*{State (stacked signal-comprised tensor)}
Once EAMs have been trained, we feed new historical prices, $s_t$, and financial news of the designated assets, to generate predictive trading signals $a^{sig}_t$. Then we stack the same new historical prices to $a^{sig}_t$ to formalize a 2-D signal-comprised tensor $s^{sc}_t$ as the data source to train SAM.
Because an SAM is connected to multiple EAMs, the 2-D signal-comprised tensors from all connected EAMs are stacked and transformed into a 3-D signal-comprised tensor called \textbf{\textit{profound state}} $v^{+}_t$, which is the state that SAM observes at each time step $t$.

\subsubsection*{Proximal policy optimization}

A PPO \cite{schulman2017proximal} agent is at the center of SAM to reallocate assets. PPO is an actor-critic style policy gradient method that has been widely used on continuous action space problems, due to its desirable performance and ease of implementation. A policy $\pi_{\theta}$ is a parametrized mapping: $S \times A \rightarrow [0,1]$ from state space to action space. Among the different objective functions of PPO, we implement the clipped surrogate objective \cite{schulman2017proximal}:
\begin{equation}
L(\theta)
=\hat{\mathbb{E}}\pi_{\theta'}[min(r_{t}(\theta)A^{\theta'}_{t},clip(r_{t}(\theta),1-\epsilon,1+\epsilon)A^{\theta'}_{t})]
\end{equation}
where \[r_{t}(\theta) = \frac{\pi_\theta(a_t|s_t)}{\pi_{\theta'}(a_{t}|s_t)}\]
and $A^{\theta'}_{t}$, the advantage function, is expressed as:
\[A^{\theta'}_{t} = Q^{\theta'}(s_t,a_t)-V^{\theta'}(s_t)\]
in which, the state-action value function $Q^{\theta'}(s_t,a_t)$ is:
\[Q^{\theta'}(s_t,a_t) = \mathbb{E}_{  \pi_{\theta'}}[\sum_{k=0}^{\infty}\gamma^{k}r_{t+k+1}\mid s_t=s,a_t=a]\]
and the value function $V^{\theta'}(s_t)$ is:
\[V^{\theta'}(s_t) = \mathbb{E}_{  \pi_{\theta'}}[\sum_{k=0}^{\infty}\gamma^{k}r_{t+k+1}\mid s_t=s]\]

For the PPO agent, we design a policy network architecture targeting the uniqueness of continuous action space in financial portfolio management problems, inspired by the EIIE topology \cite{jiang2017deep}. Because assets’ reallocated weights at time step $t$ are strictly required to total up to 1.0, we set $m^*$ normal distributions $N_1(\mu^1_t,\sigma),...,N_{m^*}(\mu^{m^*}_t,\sigma)$, and we sample $x_t \in \mathbb{R}^{m^* \times 1}$ from the distributions, where $m^*=m+1$ and $\mu_t\in \mathbb{R}^{1\times m^* \times 1}$ is the linear output of the last layer of the neural network and with standard deviation $\sigma=0$. We eventually obtain the reallocation weights $a_t=Softmax(x_t)$ and the log probability of $x_t$ for the PPO agent to learn.

\begin{figure}[!h]
        \centering
           \includegraphics[width=\linewidth]{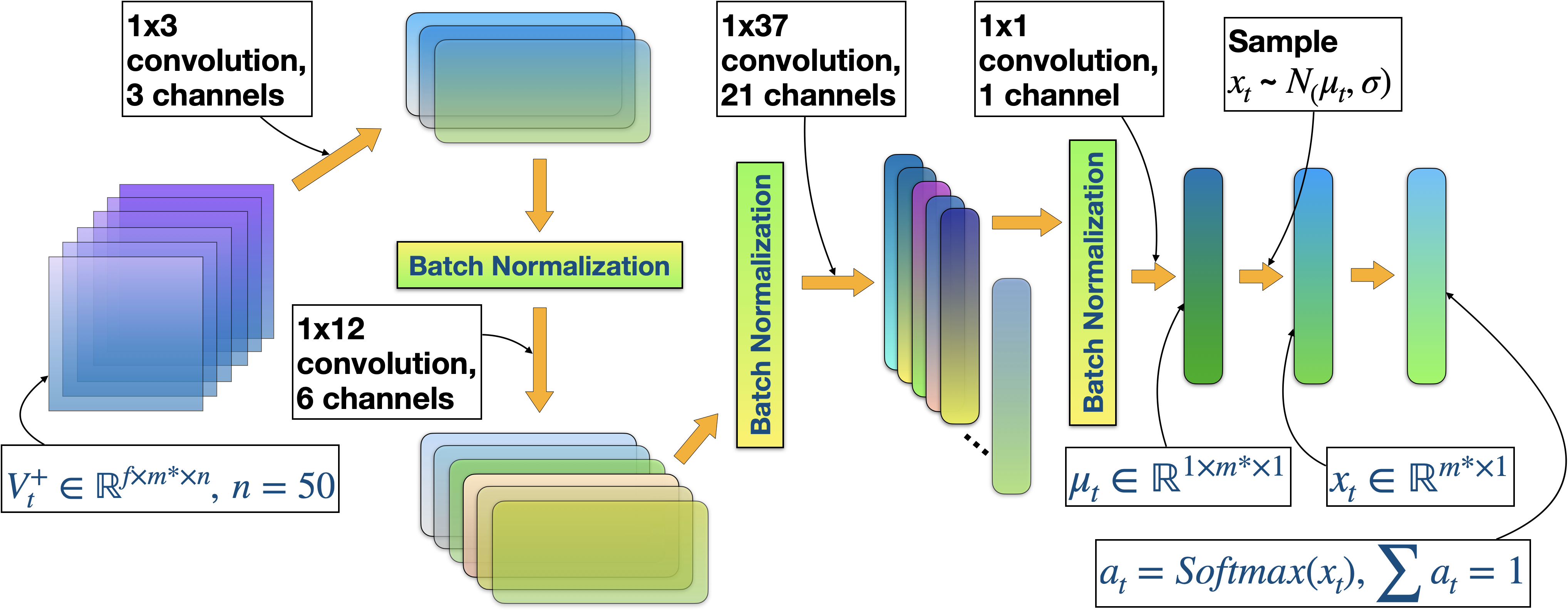}
               \caption{{\bf Policy network ($\theta'$) of SAM to accommodate PPO algorithm.} Profound state $V^+_t$ is the input of the network. $f$ is the number of features, $m^*$ is the number of assets in the portfolio, and $n=50$ is the fixed rolling-window length. After $x_t \in \mathbb{R}^{m^* \times 1}$ are sampled from the normal distributions $N_1(\mu^1_t,\sigma),...,N_{m^*}(\mu^{m^*}_t,\sigma)$, we calculate log probability of $x_t$ and obtained the reallocation weights $a_t=Softmax(x_t)$. ReLu activation function~\protect\cite{10.5555/3104322.3104425} is set after every convolutional layer, except the last one.}
          \label{fig:sam_nn}
    \end{figure}

Fig~\ref{fig:sam_nn} shows the details of the policy network (actor) of SAM, denoted by $\theta'$. Due to the resemblance and equivalence, architectures of the value network (critic) and target policy network, denoted by $\theta$, are not illustrated.

\subsubsection*{Action}
\begin{figure}[!h]
        \centering
           \includegraphics[width=0.40\linewidth]{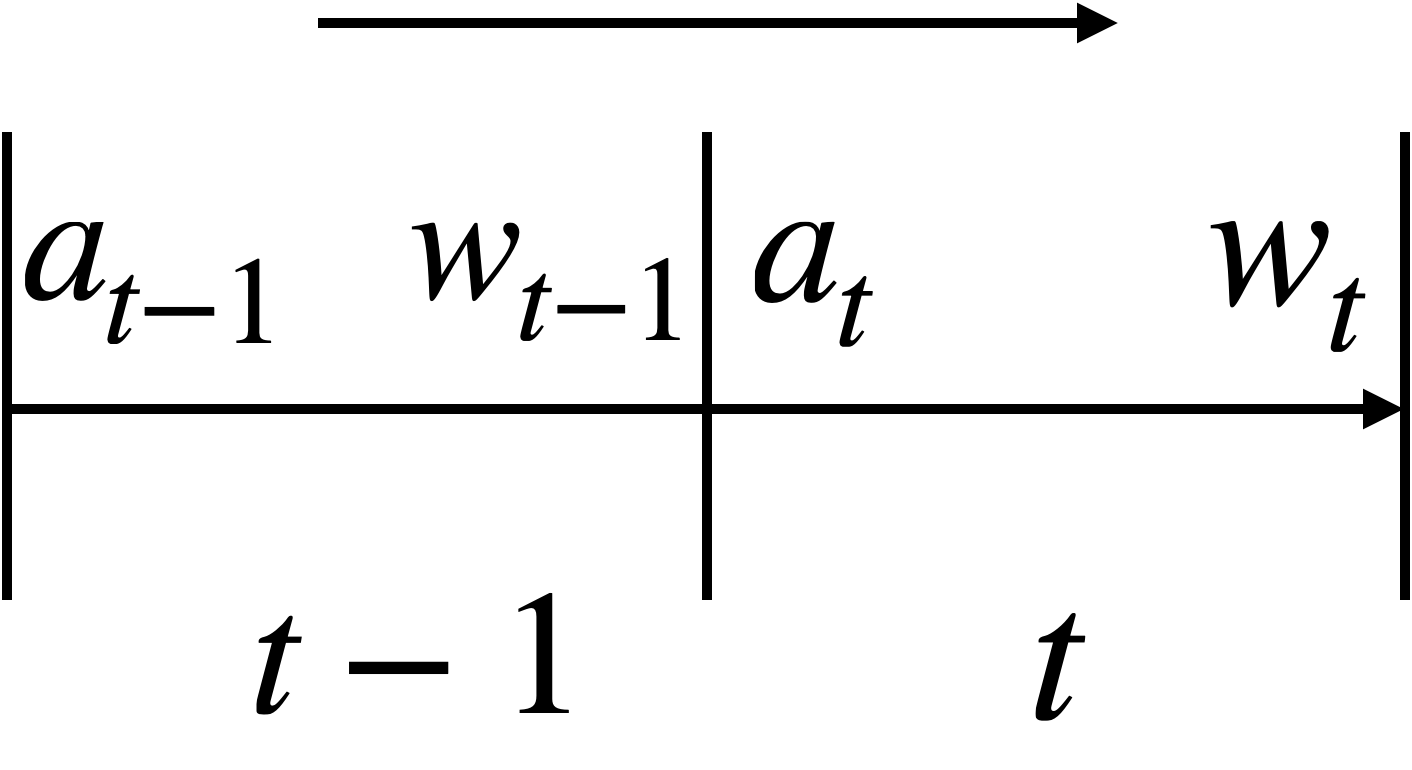}
               \caption{{\bf Transformed allocation weights due to the fluctuation in assets’ prices.}}
          \label{fig:fluctuation}
    \end{figure}

The action the PPO agent takes at each time step $t$ is
\begin{equation}
a_t = (a_{1,t},a_{2,t},...,a_{m^*,t})^T
\end{equation}
which is the vector of reallocating weights at each time step $t$, and $\sum_{i=1}^{m^*}a_{i,t}=1$. Fig~\ref{fig:fluctuation} shows the details of price fluctuations.

Once the assets are reallocated by $a_t$, the allocation weights of the portfolio eventually become
\begin{equation}
w_{t} = \frac{y_t\odot a_{t}}{y_t\cdot a_{t}}
\end{equation}
at the end of time step $t$ due to the price fluctuation during the time step period; where,
\begin{equation}
y_t = \frac{v^{+(close)}_{t}}{v^{+(close)}_{t-1}} = (1,\frac{v^{+(close)}_{2,t}}{v^{+(close)}_{2,t-1}},...,\frac{v^{+(close)}_{m^*,t}}{v^{+(close)}_{m^*,t-1}})^T
\end{equation}
is the relative price vector, that is, the changes of asset prices over time, including the prices of assets and cash. $v^{+(close)}_{i,t}$ denotes the closing price of the $i$-th asset at time $t$, where $i=\{2,...,m^*\}$, excluding cash (risk-free asset) whose closing price should always be 1.

\subsubsection*{Reward}

Inspired by \cite{jiang2017deep} in which the agent maximizes the sum of the logarithmic value, and \cite{liang2018adversarial} in which the authors try to cluster the periodic portfolio risk to alleviate the biases in training data and to prevent exposure to highly-volatile assets, we set the reward to be a risk-adjusted rate of return, $r^*_t$, which PPO agent receives at each time step $t$:
\begin{equation}
r^*_t(s_t,a_t) = \ln{(a_t\cdot y_t-\beta \sum_{i=1}^{m^*}|a_{i,t}-w_{i,t}| - \phi\sigma^{2}_{t})}
\end{equation}
where $m^*$ is the number of assets, $w_{t}$ represents the allocation weights of the assets at the end of time step $t$.
\begin{equation}
\beta \sum_{i=0}^{n}|a_{i,t}-w_{i,t}|
\end{equation}
is the transaction cost, where $\beta = 0.0025$ is the commission rate, and $\phi=0.001$ is the risk discount which can be fine-tuned as a hyperparameter.
\begin{equation}
\sigma^{2}_{t} = \frac{1}{n}\sum_{t-n+1}^{t}\sum_{i=1}^{m^*}(y_{i,t-n+1},\overline{y_{i,t-n+1}})^2
\end{equation}
measuring the volatility of fluctuation in assets’ prices during the last $n$ days.
\begin{equation}\overline{y_{i,t-n+1}}=\frac{1}{n}\sum_{t-n+1}^{t}y_{i,t-n+1}
\end{equation}
is the volatility of the profit of an individual asset. We expect the agent to secure a maximum risk-adjusted rate of return (capital gain) every time step, as what is expected from human portfolio managers.

\section*{Experiments}
In this section, we build different portfolios, and train MSPM to periodically reallocate the assets in each portfolio. The portfolios, datasets, and performance metrics for benchmarking will be introduced and described. After that, we explain and discuss the experimental results and examine MSPM's stability of daily rate of return. We also inspect the signal generation and position-holding of EAMs. In the end, we validate the necessity of EAM by back-testing four different portfolios. The back-testing performance of MSPM will be compared with the existing baselines.

\subsection*{Preliminaries}

\subsubsection*{Portfolios}
We first propose two portfolios: (a) and (b) to compare back-testing performance. Portfolio(a) includes three stocks: Apple, AMD, and Alphabet (symbol codes: [AAPL, AMD, GOOGL]), and Portfolio(b) includes three other stocks: Alphabet, Nvidia, and Tesla (symbol codes: [GOOGL, NVDA, TSLA]). To build portfolio(a) and portfolio(b), we trained two SAM/MSPMs: SAM/MSPM(a) and SAM/MSPM(b). Additionally, the two SAMs shared the same EAM for the stock in common: Alphabet (GOOGL). Later, we propose two other portfolios (c) and (d), which make four portfolios in total, to validate the necessity of EAM. Details can be found in the Validation of EAM section. For all these four portfolios, we set initial portfolio value to be $p_0=10,000$.

\subsubsection*{Data ranges}
Among the EAMs to be trained, the foundational EAM (AAPL) is trained initially, and its parameters are shared with other EAMs as their foundation for transfer learning. As shown in Table~\ref{tab:datasets}, EAM-training data, ranging from January 2009 to December 2015, contains the historical prices ($s_t$) and news sentiments ($\rho_t$) of the stocks, including AAPL, in portfolios (a) and (b). EAM-predicting data, with the same data structure as EAM-training and ranging from January 2016 to December 2020, is used for EAMs to predict and generate trading signals (actions of DQN agents). Then, EAM-predicting data along with the generated trading signals became the signal-comprised data for SAM/MSPMs. There are three datasets of signal-comprised data: SAM/MSPM-training and SAM/MSPM-validating to train and validate SAMs, respectively; and SAM/MSPM-experiment, from January 2020 to December 2020, for back-testing and other experiments. Details can be found in Table~\ref{tab:datasets}. It is worth noting that a low percentage (9.51\%) of missing values from the alternative data (sentiments) shall not affect MSPM's scalability nor reusability since, as a general framework, MSPM is neutral on the structures, types, or sources of the data input.

\begin{table}[!ht]
\centering
\caption{
{\bf Date ranges of the data}}
\begin{tabular}{@{}lrrl@{}}
\hline\noalign{\smallskip}
Purpose & Data Range \\
\noalign{\smallskip}\hline\noalign{\smallskip}
EAM-training & Jan 2009$\sim$Dec 2015 \\
EAM-predicting & Jan 2016$\sim$Dec 2020 \\
\noalign{\smallskip}\hline\noalign{\smallskip}
SAM/MSPM-training & Jan 2016$\sim$Dec 2018 \\
SAM/MSPM-validation & Jan 2019$\sim$Dec 2019 \\
SAM/MSPM-experiment & Jan 2020$\sim$Dec 2020 \\
\noalign{\smallskip}\hline
\end{tabular}
\begin{flushleft} EAM-training dataset includes the historical prices ($s_t$) and news sentiments ($\rho_t$) of all assets in the portfolios constructed, and is used to train the AAPL-based foundational EAM, and transfer learning for the four other assets. EAM-predicting dataset includes new historical prices ($s_t$) and news sentiments ($\rho^*_t$) for EAMs to generate signal-comprised tensors ($s^{sc}_t=(s_t,a^{sig}_t)$) to formalize the SAM/MSPM-training ($v^+$) data. SAM/MSPM-validation and SAM-back-testing data have the same structure as SAM/MSPM-training but are used solely for the purposes of validation and back-testing.
\end{flushleft}
\label{tab:datasets}
\end{table}

\subsubsection*{Performance metrics}
We use the following performance metrics to measure the performances of the baselines and MSPM system.

\begin{itemize}
\item{\textbf{Daily Rate of Return (DRR)}}
\begin{equation}
DRR_T = \frac{1}{T}\sum_{t=1}^{T}\exp(R_t),
\end{equation}
where $T$ is the terminal time step, and \begin{equation}
R_t = \ln{(a_t\cdot y_t-\beta \sum_{i=1}^{m^*}|a_{i,t}-w_{i,t}|)}
\end{equation}
is the risk-unadjusted periodic (daily) rate of return obtained at every time step, where $\beta \sum_{i=1}^{m^*}|a_{i,t}-w_{i,t}|$ is the transaction cost and $\beta$ = 0.0025 is the commission rate.

\item{\textbf{Accumulated rate of return (ARR)}}
The accumulated rate of return (ARR) \cite{doi:10.1080/14697688.2011.570368} is
\begin{equation}
ARR_T = \frac{p_T}{p_0},
\end{equation}
where $T$ is the terminal time step, $p_0$ is the portfolio value at the initial time step, and
\begin{equation}
p_T = p_0 \exp{(\sum_{t=1}^{T}R_t)}
\end{equation}
which stands for the portfolio value at the terminal time step.

\item{\textbf{Sortino ratio (SR)}}
Sortino ratio \cite{Sortino59} is often referred to as a risk-adjusted return, which measures the portfolio performance compared to a risk-free return, adjusted by the portfolio’s downside risk. In our case, Sortino ratio is calculated as
\begin{equation}
SR = \frac{\frac{1}{T}\sum_{t=1}^{T}\exp(R_t)-R_f}{\sigma^{downside}}
\end{equation}
where $R_t$ is the risk-unadjusted periodic (daily) rate of return.  Portfolio's downside risk $\sigma^{downside}$ is calculated as
\begin{equation}
\sigma^{downside} = \sqrt{\textnormal{Var}(R_l-R_f)},
\end{equation}
where $R_f$ is the risk-free return and conventionally equals zero, $R_l$ are the less-than-zero returns in $R_t$ for all $t$, and $t=T$ is the terminal time step.

\item{\textbf{Max drawdown (MD)}}
MD is the biggest drop (in \%) between the highest (peak) and lowest (valley) of the accumulated rate of return of a certain period of time.

\end{itemize}

For DRR, ARR and SR, we want them to be as high as possible, whereas we want MD to be as low as possible.

\subsection*{Results and discussion}

\subsubsection*{Back-testing performance}

We back-test and compare the performance of our MSPM system to different baselines, including the traditional and cuttings-edge RL-based portfolio management strategies~\cite{10.1145/2512962,mlfinlab}. The baselines are listed as follows:

\begin{itemize}
\item{\textbf{CRP}} stands for (Uniform) Constant Rebalanced Portfolio, which involves investing an equal proportion of capital in each asset, namely 1/N, which seems simple but, in fact, challenging to beat \cite{10.1093/rfs/hhm075}.

\item{\textbf{Buy and hold (BAH)}} strategy involves investing without rebalancing. Once the capital is invested, no further allocation will be made.

\item{\textbf{Exponential gradient portfolio (EG)}} strategy involves investing capital into the latest stock with the best performance and uses a regularization term to maintain the portfolio information.

\item{\textbf{Follow the regularized leader (FTRL)}} strategy tracks the Best Constant Rebalanced Portfolio until the previous period, with an additional regularization term. This strategy reweights based on the entire history of the data with an expectation to obtain maximum returns.

\item{\textbf{ARL}} refers to the adversarial deep reinforcement learning in portfolio management (Adversarial PG)~\cite{liang2018adversarial}, which is a state-of-the-art (SOTA) RL-based portfolio management method.

\end{itemize}

\begin{figure}[!h]
        \centering
           \includegraphics[width=0.8\linewidth]{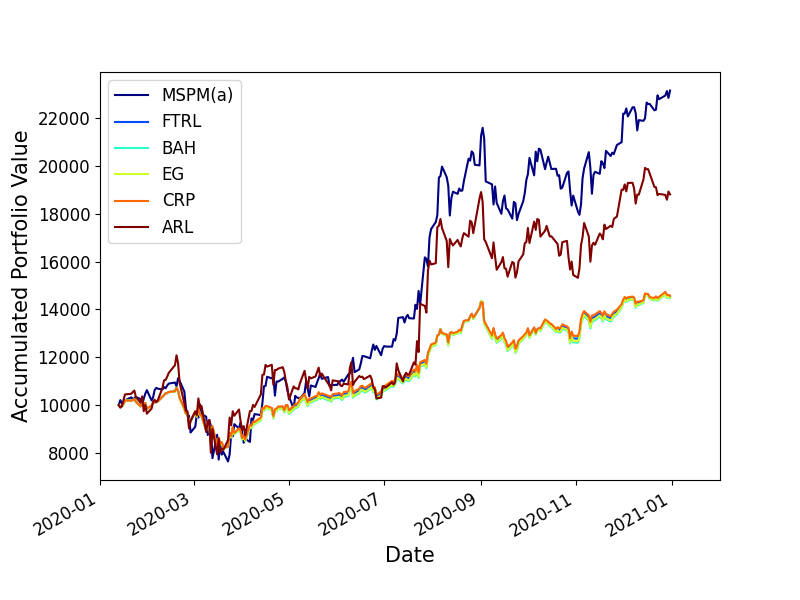}
               \caption{{\bf MSPM(a) outperforms all baselines on Portfolio(a) in terms of the accumulated portfolio value in back-testing.}}
          \label{fig:pa}
\end{figure}

\begin{figure}[!h]
        \centering
           \includegraphics[width=0.8\linewidth]{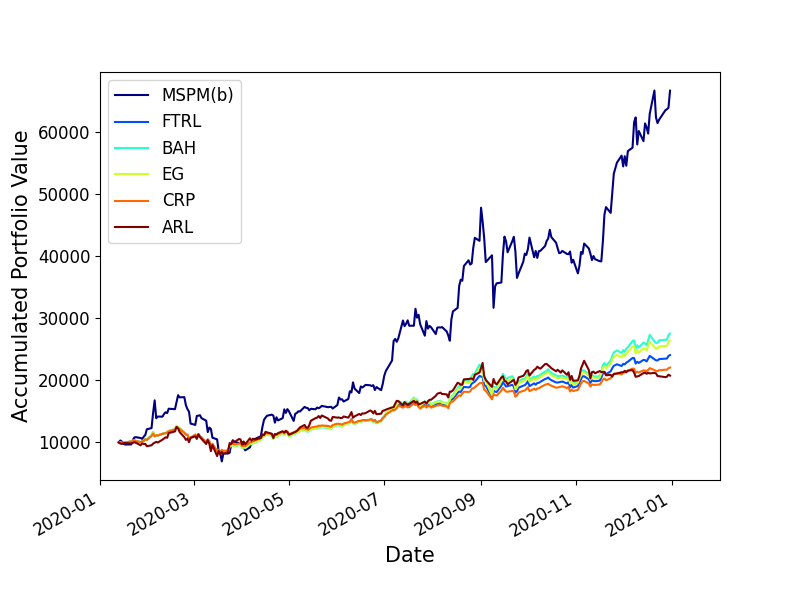}
               \caption{{\bf MSPM(b) outperforms all baselines on Portfolio(b) in terms of the accumulated portfolio value in back-testing.}}
          \label{fig:pb}
\end{figure}

As shown in Fig~\ref{fig:pa} and~\ref{fig:pb}, for both portfolios (a) and (b), MSPM system improves ARR, by at least 49.3\% and 426.6\% compared to ARL, a SOTA RL-based PM method, and by 186.5\% and 369.8\% compared to CRP, a traditional PM strategy, during the year of 2020. The result demonstrates the advantage of MSPM at gaining capital returns. Table~\ref{tab:backtest} gives details about MSPM's outperformance over existing baselines in terms of the ARR and DRR. Further, MSPM's superior performance on SR indicates that MSPM takes better consideration of harmful volatility and achieves higher risk-adjusted returns.\\
 
\begin{table}[!ht]
\begin{adjustwidth}{-2.25in}{0in}
\centering
\caption{\bf Comparison of back-testing performance of the baselines and MSPM}
\begin{tabular}{|c|c|c|c|c|c|c|c|c|c|c|c|c|}
\hline
       & \multicolumn{6}{c|}{Portfolio (a)} & \multicolumn{6}{c|}{Portfolio (b)} \\ \hline
Metric & CRP & BAH & EG & FTRL & ARL & MSPM & CRP & BAH & EG & FTRL & ARL & MSPM \\ \hline
DRR (\%) & 0.175          & 0.173 & 0.173 & 0.174 & 0.333 & \textbf{0.404} & 0.350          & 0.460 & 0.440 & 0.395 & 0.367 & \textbf{0.938} \\ \hline
ARR (\%) & 45.9           & 44.7  & 44.9  & 45.4  & 88.1  & \textbf{131.5} & 120.6          & 175.4 & 164.0 & 140.7 & 107.6 & \textbf{566.6} \\ \hline
MD (\%)  & \textbf{-23.3} & -23.6 & -23.5 & -23.4 & -34.3 & -31.3          & \textbf{-33.6} & -35.7 & -35.3 & -34.5 & -37.6 & -60.6          \\ \hline
SR       & 1.95           & 1.88  & 1.89  & 1.92  & 2.13  & \textbf{2.86}  & 3.24           & 3.54  & 3.50  & 3.38  & 2.35  & \textbf{4.18}  \\ \hline
\end{tabular}
\label{tab:backtest}
\end{adjustwidth}
\end{table}

It is worth noting that for portfolio (a), both MSPM and ARL achieve promising SR, but for portfolio (b), only MSPM has a much better Sortino ratio than ARL, which indicate MSPM's higher adaptability to the ever-changing market compared to not only the traditional strategies but also the preceding RL-based method.

\subsubsection*{Stability of daily rate of return (DRR)}
Due to the high max drawdown (MD) of MSPM for portfolio(b) (60.6\%), we want to examine and compare the general stability of DRR between MSPM and the state-of-the-art RL-based method: ARL. For this purpose, we first calculate DRR's 5-day rolling standard deviation (RstdDRR) as the proxy of the stability of DRR. Higher RstdDRR indicates lower stability of DRR.\\

To calculate the RstdDRR, we first calculate the simple moving average (SMA)~\cite{ADictionaryofFinanceandBanking} of DRR$\in\mathbb{R}^{k}$ for the past n data-points (days) by the following formula:

\begin{equation}
SMA_{i}=\frac{DRR_{i-n+1}+DRR_{i-n+2}+\ldots+DRR_{i}}{n}
\end{equation}

for $i=n,...,k$. Then, we subtract $SAM_{i}$ from the 5-day DRRs used in the calculation, and then take the square root of the squared summation to have the rolling standard deviation: RstdDRR$\in\mathbb{R}^{k-n}$:

\begin{equation}
RstdDRR_{i}=\sqrt{\frac{\left(DRR_{i-n+1}-SMA_{i}\right)^{2}+\left(DRR_{i-n+2}-SMA_{i}\right)^{2}+\ldots+(DRR_{i}-SMA_{i})^{2}}{n}}
\end{equation}
where $i=n,...,k$.

\begin{figure}[!h]
 \includegraphics[width=\linewidth]{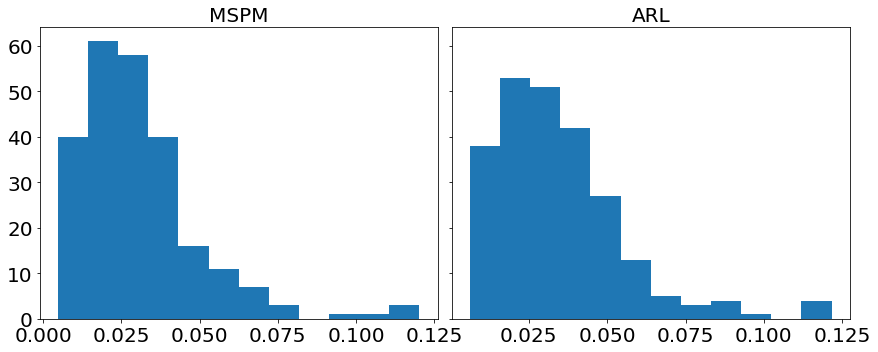}
\caption{{\bf For portfolio(a), histograms of MSPM and ARL's 5-day RstdDRR depict right-skewed distributions.}}
\label{fig:rstdhist_a}
\end{figure}

\begin{figure}[!h]
 \includegraphics[width=\linewidth]{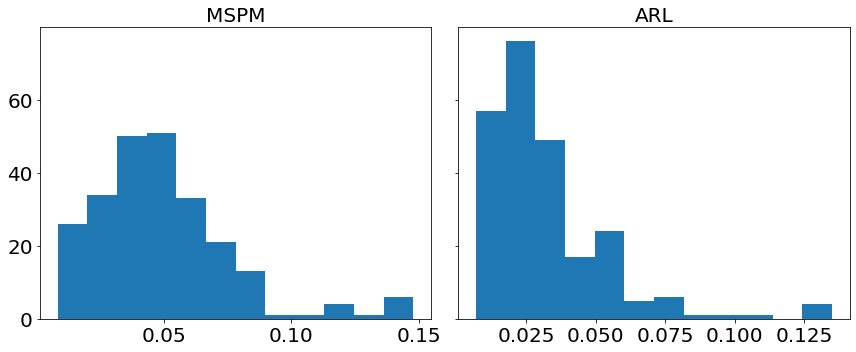}
\caption{{\bf For portfolio(b), histograms of MSPM and ARL's 5-day RstdDRR depict right-skewed distributions.}}
\label{fig:rstdhist_b}
\end{figure}

Fig~\ref{fig:rstdhist_a} shows the histograms of MSPM and ARL's RstdDRR for portfolio(a), and histograms in Fig~\ref{fig:rstdhist_b} are for portfolio(b). According to Fig~\ref{fig:rstdhist_a}, the right tail of ARL's RstdDRR is fatter than that of MSPM's RstdDRR, and MSPM has a lower average RstdDRR ($M_{(a)}=0.031$, $SD_{a}=0.019$) than ARL ($M_{(a)}=0.034$, $SD_{a}=0.020$), indicating MSPM has higher stability of DRR on portfolio(a). However, Fig~\ref{fig:rstdhist_b} depicts that the right tail of MSPM's RstdDRR is fatter than that of ARL's RstdDRR, and the mean of MSPM's RstdDRR ($M_{(b)}=0.049$, $SD_{b}=0.027$) is larger than the mean of ARL's RstdDRR ($M_{(b)}=0.032$, $SD_{b}=0.022$). For more information, figures in~\nameref{S2_Fig} and~\nameref{S3_Fig} give the comparison between MSPM and ARL's RstdDRR for portfolio (a) and (b). As shown in~\nameref{S2_Fig}, the RstdDRR of MSPM is less volatile than that of ARL, but it is the opposite case in~\nameref{S3_Fig}.\\

Since the histograms in Fig~\ref{fig:rstdhist_a} and~\ref{fig:rstdhist_b} show skewed bell shapes, we use Shapiro-Wilk test~\cite{10.1093/biomet/52.3-4.591} to confirm the normality of the distributions. After that, we use Levene's test~\cite{levene} to examine the variance equality. We use Python's SciPy library to perform these two tests. By implementing Shapiro–Wilk test, we find that MSPM and ARL's RstdDRR are not statistically from normal distributions for both portfolios (p-values are less than 0.05). Moreover, according to Levene's test, MSPM and ARL's RstdDRR do not always have homogeneity of variance: for portfolio (a) they do, whereas for portfolio(b) they do not. With the assumptions verified, we perform the one-tail and two-sample Mann–Whitney U test~\cite{10.2307/2236101} (a non-parametric version of unpaired t-test) to rigorously compare MSPM and ARL's stability of DRR, also using Python's SciPy library. For portfolio(a), because the mean RstDRR of MSPM is less than the mean RstDRR of ARL, the hypothesis $H_{0}$ is that MSPM has a lower or same stability than ARL (the group mean of RstdDRR of MSPM is greater or equal to that of ARL), and the alternative hypothesis $H_{a}$ is that MSPM has higher stability than ARL(the group mean of RstdDRR of MSPM is less than that of ARL). For portfolio(b), because the mean RstDRR of MSPM is higher than the mean RstDRR of ARL, the hypothesis $H_{0}$ is that MSPM has higher or same stability than ARL (the group mean of RstdDRR of MSPM is less or equal to that of ARL), and the alternative hypothesis $H_{a}$ is that MSPM has a lower stability than ARL(the group mean of RstdDRR of MSPM is greater than that of ARL). We set the significance level to be .05. If the p-value from the test is less than 0.05, we reject $H_{0}$ and accept $H_{a}$; otherwise, we accept the null hypothesis $H_{0}$. The detailed settings of the statistical test are:

\begin{itemize}
\item{Statistical test: one-tail and two-sample Mann–Whitney U test}
\item{For portfolio (a), null hypothesis $H^{Pa}_{0}: \mu^{RstdDRR}_{MSPM} - \mu^{RstdDRR}_{ARL} \geq 0$}
\item{For portfolio (a), alternative hypothesis $H^{Pa}_{a}: \mu^{RstdDRR}_{MSPM} - \mu^{RstdDRR}_{ARL} < 0$}
\item{For portfolio (b), null hypothesis $H^{Pb}_{0}: \mu^{RstdDRR}_{MSPM} - \mu^{RstdDRR}_{ARL} \leq 0$}
\item{For portfolio (b), alternative hypothesis $H^{Pb}_{a}: \mu^{RstdDRR}_{MSPM} - \mu^{RstdDRR}_{ARL} > 0$}
\item{Significance level: .05}
\end{itemize}

As the results represented in Table~\ref{tab:postDRR}, MSPM has significantly higher stability of DRR than ARL for portfolio(a) by rejecting $H_{0}$ and accepting $H_{a}$ ($U_{a}=25426.0$, $p-value=.005$). For portfolio(b), because $H_{0}$ is accepted ($U_{b}=16209.0$, $p-value<.001$), we confirm that MSPM has lower stability of DRR than ARL. The conclusions are aligned with the MD in Table~\ref{tab:backtest} and the underwater plots in~\nameref{S4_Fig},~\nameref{S5_Fig},~\nameref{S6_Fig} and~\nameref{S7_Fig} which illustrate the drawdowns during year 2020. It is clear in ~\nameref{S4_Fig} and~\nameref{S5_Fig} that ARL has more frequent and intensive drawdowns for portfolio(a) compared to MSPM, but MSPM becomes the more volatile one for portfolio(b) according to~\nameref{S6_Fig} and~\nameref{S7_Fig}. The results indicate that although MSPM achieves an outstanding performance in gaining capital returns, it does not naturally come with higher stability. However, low stability (or high risk) does not necessarily refer to danger. Since for both portfolio (a) and (b), MSPM has the highest Sortino ratios, which consider only the downside risk, MSPM's lower stability for portfolio (b) may come from a higher upside risk. In conclusion, there should be a trade-off between performance and stability, and this can be further investigated and considered in future studies.

\begin{table}[!ht]
\begin{adjustwidth}{-2.25in}{0in}
\centering
\caption{{\bf Results of statistical test on the RstdDRR of MSPM and ARL}}
\begin{tabular}{cccccccc}
\hline
\multicolumn{1}{c|}{} &
  \multicolumn{2}{c|}{MSPM (n=241)} &
  \multicolumn{2}{c|}{ARL (n=241)} &
  \multicolumn{1}{c|}{} &
  \multicolumn{1}{c|}{} &
   \\ \hline
\multicolumn{1}{c|}{Portfolio} &
  \multicolumn{1}{c|}{M(SD)} &
  \multicolumn{1}{c|}{Normality} &
  \multicolumn{1}{c|}{M(SD)} &
  \multicolumn{1}{c|}{Normality} &
  \multicolumn{1}{c|}{EV} &
  \multicolumn{1}{c|}{U} &
  p-value \\ \hline
\multicolumn{1}{c|}{(a)} &
  \multicolumn{1}{c|}{0.031(0.019)} &
  \multicolumn{1}{c|}{$6.98*10^{-15}$} &
  \multicolumn{1}{c|}{0.034(0.020)} &
  \multicolumn{1}{c|}{$3.39*10^{-13}$} &
  \multicolumn{1}{c|}{0.324} &
  \multicolumn{1}{c|}{25426.0} &
  .005 \\ \hline
\multicolumn{1}{c|}{(b)} &
  \multicolumn{1}{c|}{0.049(0.027)} &
  \multicolumn{1}{c|}{$1.48*10^{-11}$} &
  \multicolumn{1}{c|}{0.032(0.022)} &
  \multicolumn{1}{c|}{$1.19*10^{-16}$} &
  \multicolumn{1}{c|}{10.070} &
  \multicolumn{1}{c|}{16209.0} &
  $<.001$ \\ \hline
\multicolumn{8}{l}{M: Mean; SD: Standard deviation; Normality: Shapiro–Wilk test; EV: Levene's test; U: U-Statistics} \\ \hline
\end{tabular}
\label{tab:postDRR}
\end{adjustwidth}
\end{table}

\subsubsection*{EAM: Case study}
\begin{figure}[!h]
 \includegraphics[width=\linewidth]{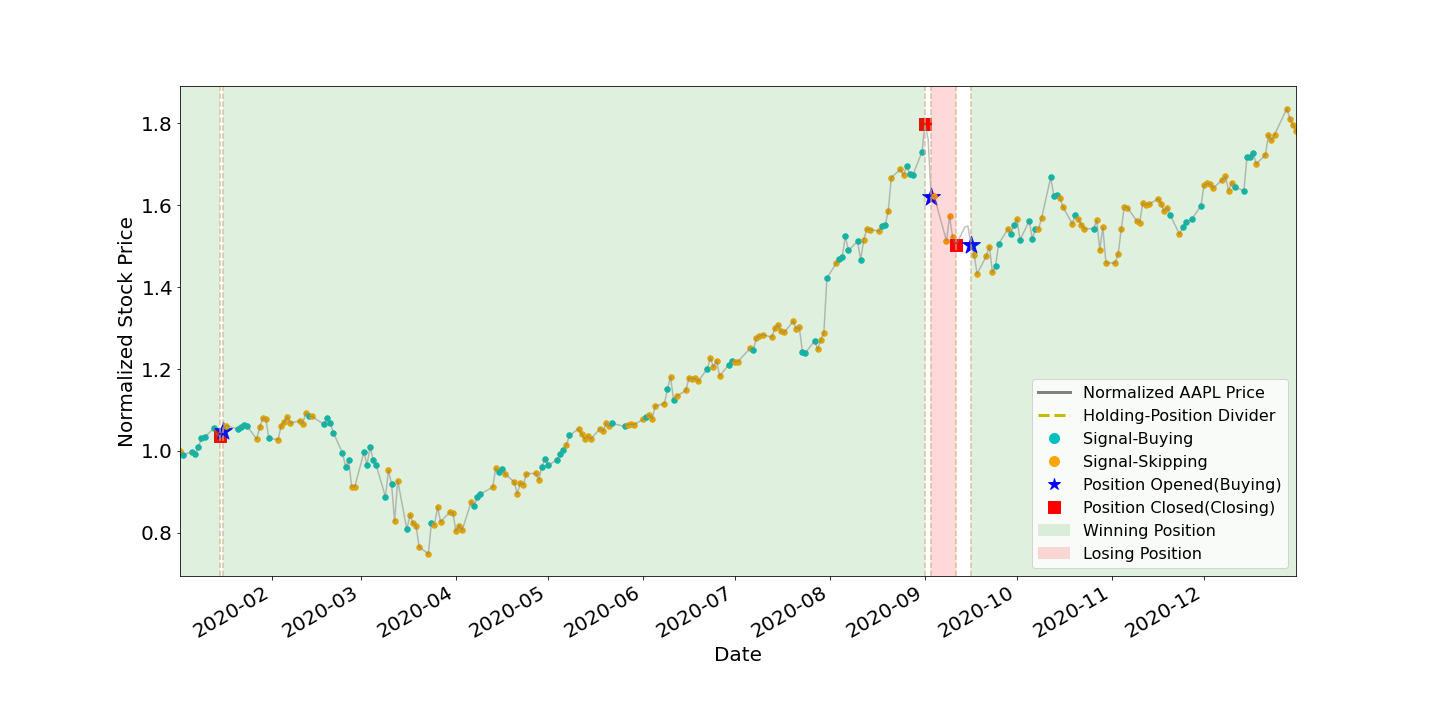}
\caption{{\bf Signals and position-holding of AAPL's EAM.}}
\label{fig:cs_aapl}
\end{figure}

\begin{figure}[!h]
 \includegraphics[width=\linewidth]{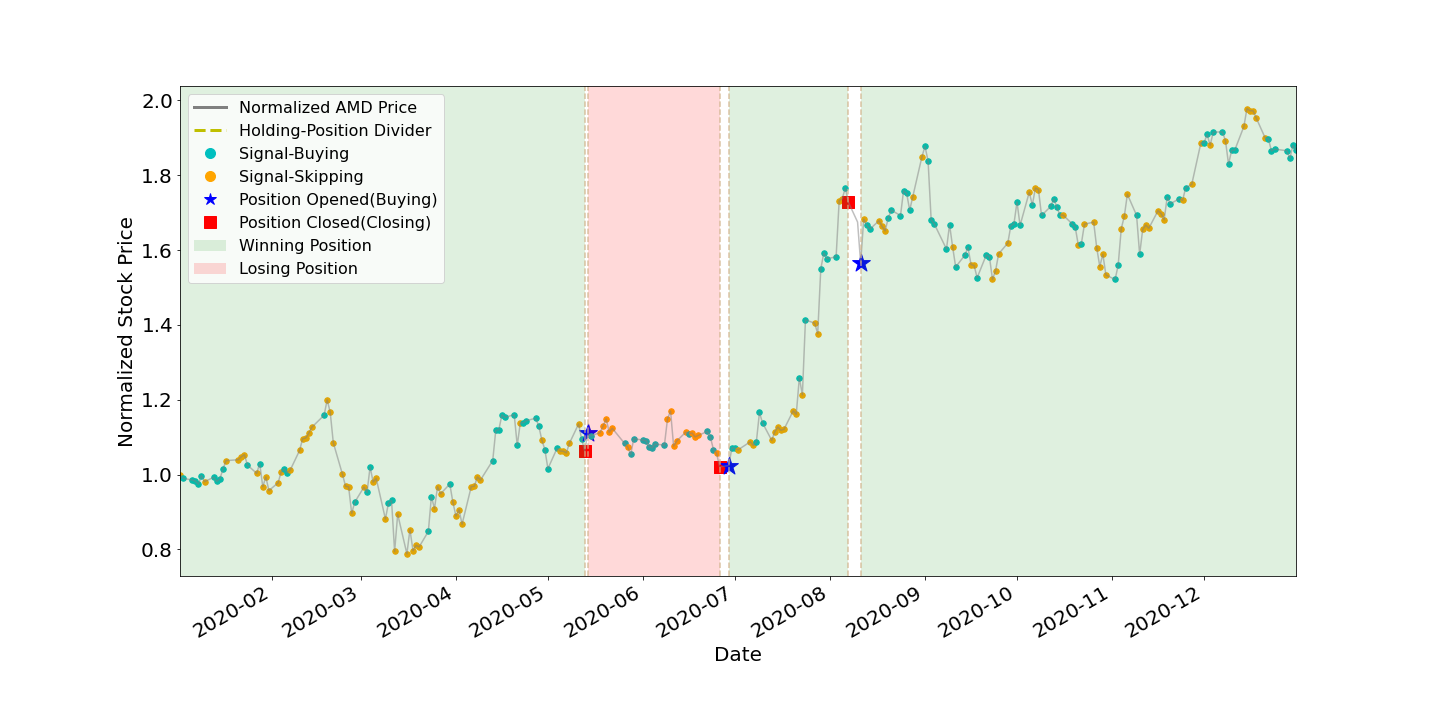}
\caption{{\bf Signals and position-holding of AMD's EAM.}}
\label{fig:cs_amd}
\end{figure}

\begin{figure}[!h]
 \includegraphics[width=\linewidth]{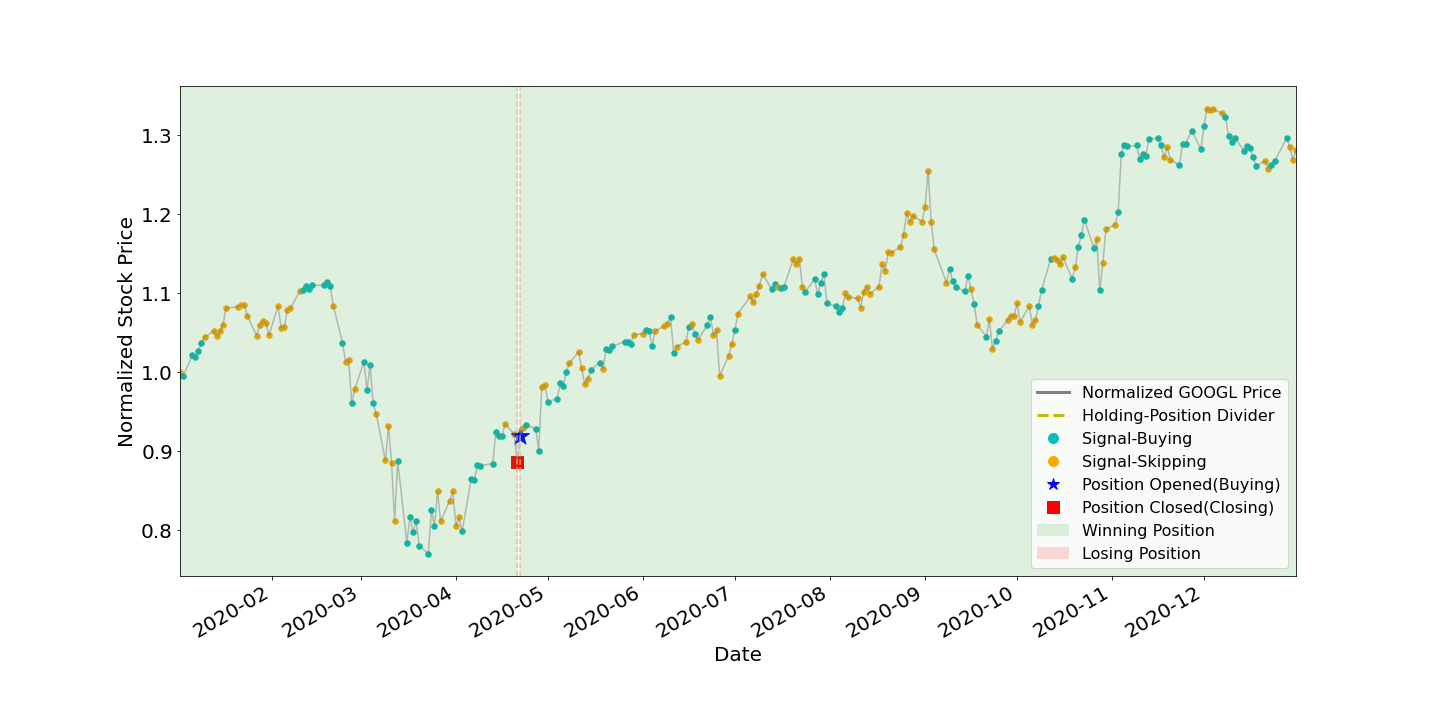}
\caption{{\bf Signals and position-holding of GOOGL's EAM.}}
\label{fig:cs_googl}
\end{figure}

\begin{figure}[!h]
 \includegraphics[width=\linewidth]{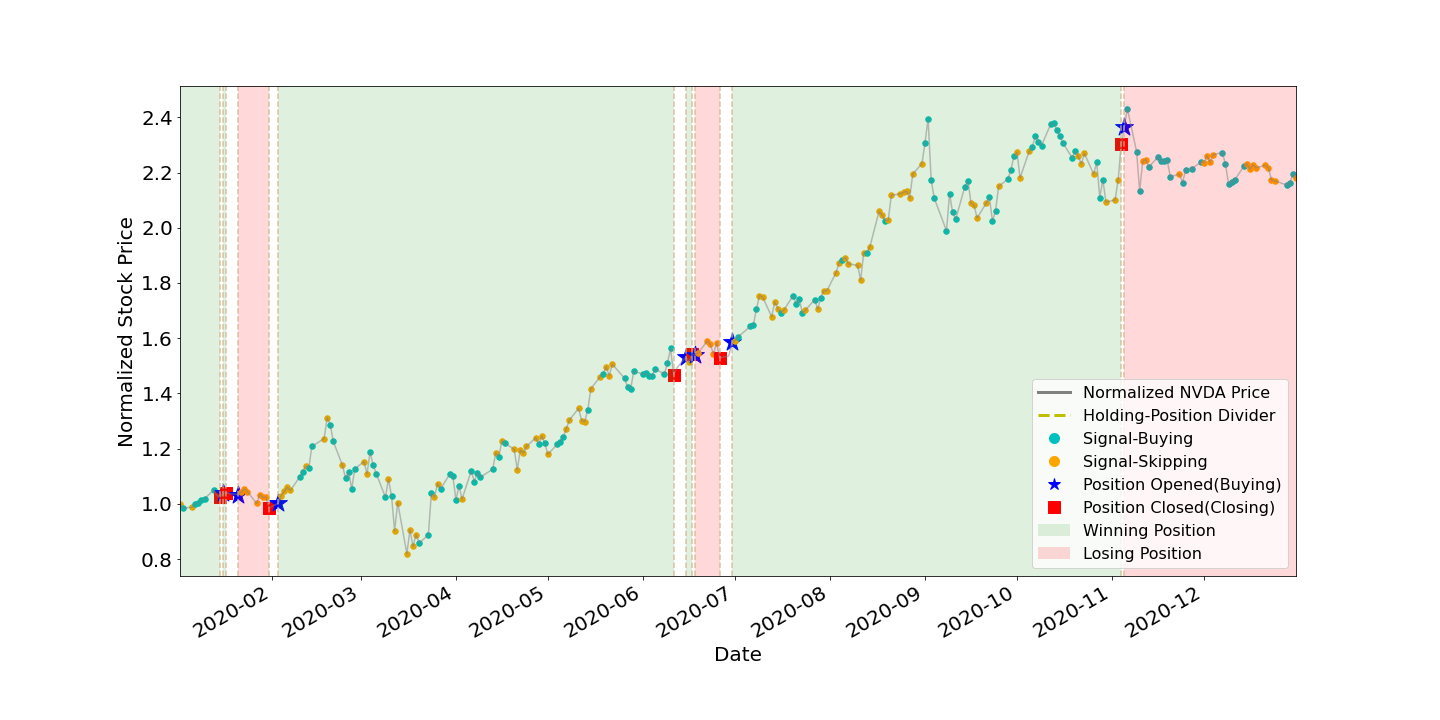}
\caption{{\bf Signals and position-holding of NVDA's EAM.}}
\label{fig:cs_nvda}
\end{figure}

\begin{figure}[!h]
 \includegraphics[width=\linewidth]{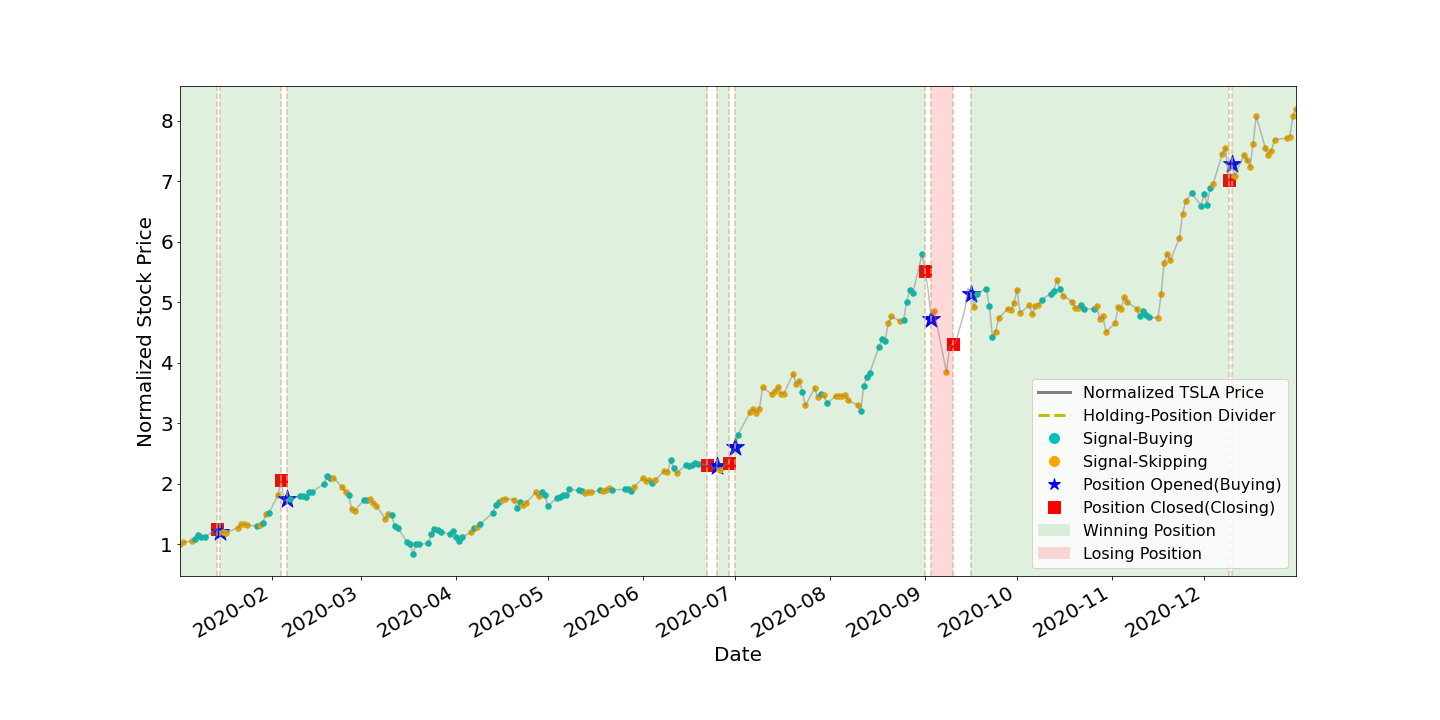}
\caption{{\bf Signals and position-holding of TSLA's EAM.}}
\label{fig:cs_tsla}
\end{figure}

To better understand how EAM contributes to SAM, we illustrate the position-holding information using the signals generated by the EAMs of portfolio (a) and (b) in Fig~\ref{fig:cs_aapl},~\ref{fig:cs_amd},~\ref{fig:cs_googl},~\ref{fig:cs_nvda}, and~\ref{fig:cs_tsla}. The figures represent the five underlying assets: AAPL, AMD, GOOGL, NVDA, TSLA. In each plot, signals of Buying and Skipping are marked with cyan and orange circles, and the positions opened or closed are marked with either star or square symbols. The grey line is the normalized price movement. A position is opened when the first Buying signal is generated after the latest position has been closed. A position is closed when the first Closing signal is generated after a position has been opened and not yet been closed. We use dashed lines to divide different position-holding periods. If a position is profit-making based on the opening and closing prices, we color the period as light green (winning position), otherwise light red. Period of no-position will be left as blank. According to the results illustrated in the figures, the positions are opened and closed at just the right timings by the corresponding EAMs for most assets.

As shown in Table~\ref{tab:cs_EAM}, the number of positions opened by any EAM is less than ten, and the highest is NVDA and TSLA's eight opened positions. The most profit-making EAM is TSLA, with ARR of 799\%. These results exemplify the high quality and reliability of the signals generated by the EAMs. The winning rates of all the five EAMs are more than 50\%. Since averaged winning rate is 80\%, it indicates that even with a mediocre averaged winning rate, SAM still can efficiently utilize the information generated by the EAMs and has the outperformance compared to ARL. The results also indicate that the MSPM can perform even better if we improve the winning rate of EAMs.

\begin{table}[!ht]
\begin{adjustwidth}{-2.25in}{0in}
\centering
\caption{{\bf Statistics of EAMs' position-holding during year 2020}}
\begin{tabular}{cccc|c|c}
\hline
\multicolumn{1}{c|}{Asset/EAM} & \multicolumn{1}{c|}{\# of positions} & \multicolumn{1}{c|}{\# of winning positions} & \# of losing positions & Wining rate (\%) & ARR (\%) \\ \hline
\multicolumn{1}{c|}{AAPL}  & \multicolumn{1}{c|}{4} & \multicolumn{1}{c|}{3} & 1 & 75\%           & 124\% \\ \hline
\multicolumn{1}{c|}{AMD}   & \multicolumn{1}{c|}{4} & \multicolumn{1}{c|}{3} & 1 & 75\%           & 125\% \\ \hline
\multicolumn{1}{c|}{GOOGL} & \multicolumn{1}{c|}{2} & \multicolumn{1}{c|}{2} & 0 & 100\%          & 44\%  \\ \hline
\multicolumn{1}{c|}{NVDA}  & \multicolumn{1}{c|}{8} & \multicolumn{1}{c|}{5} & 3 & 62\%           & 115\% \\ \hline
\multicolumn{1}{c|}{TSLA}  & \multicolumn{1}{c|}{8} & \multicolumn{1}{c|}{7} & 1 & 88\%           & 799\% \\ \hline
                           &                        &                        &   & Averaged: 80\% &       \\ \hline
\end{tabular}
\label{tab:cs_EAM}
\end{adjustwidth}
\end{table}

\subsubsection*{Validation of EAM}
\begin{figure}[!h]
        \centering
           \includegraphics[width=\linewidth]{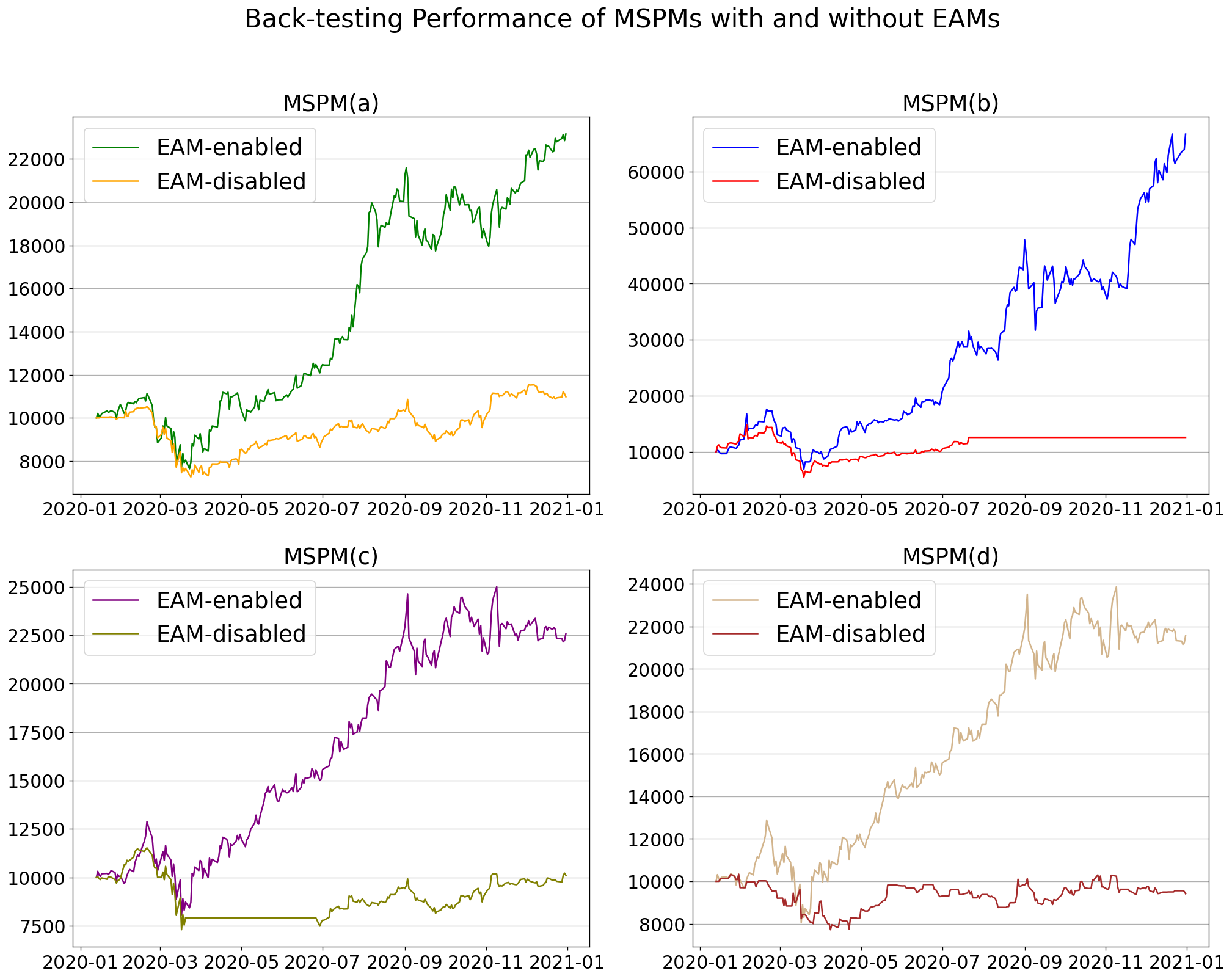}
               \caption{{\bf Accumulated portfolio values of MSPMs, with and without EAMs, from back-testing for portfolio (a), (b), (c) and (d).} For all the four portfolios, EAM-enabled MSPMs perform significantly better than EAM-disabled MSPMs.}
          \label{fig:eam_val}
\end{figure}

As EAMs provide the trading signal-comprised information to SAMs, we intend to verify the indispensability of EAM by comparing the performance of MSPMs with and without EAMs. For this purpose, we set four different portfolios: (a), (b), (c), and (d), in which (c) and (d) are newly introduced. Portfolio(c) consists of three stocks: Alphabet, Nvidia, and Amazon (symbol codes: [GOOGL, NVDA, AMZN]), and portfolio(d) consists of three other stocks: Nvidia, Facebook, and Microsoft (symbol codes: [NVDA, FB, MSFT]). Two MSPMs/SAMs share the same EAM for the common stocks, which are NVDA and AMZN. The initial portfolio values are still set to be $10,000$. Fig~\ref{fig:eam_val} shows EAM-enabled and EAM-disabled MSPMs' the accumulated returns of different portfolios. As shown in the figure, EAM-enabled MSPMs always perform better than EAM-disabled MSPMs, and this conclusion can be reconfirmed by Table~\ref{tab:EAM_backtest}. As listed in the table, EAM-enabled MSPMs largely outperform EAM-disabled MSPMs in terms of DRR, ARR, and SR. In terms of portfolio (d), EAM-enabled MSPM achieves ARR and SR of 115.6\% and 2.45, whereas EAM-disabled MSPM's ARR and SR is -5.9\% and 0.01. The results validate that the SAMs can only have an ideal performance with the trading signal-comprised information from EAMs.

\begin{table}[!ht]
\begin{adjustwidth}{-2.25in}{0in}
\centering
\caption{\bf Comparison of back-testing performance of EAM-enabled and EAM-disabled MSPMs}
\begin{tabular}{|c|c|c|c|c|c|c|c|c|}
\hline
 & \multicolumn{2}{c|}{Portfolio (a)} & \multicolumn{2}{c|}{Portfolio (b)} & \multicolumn{2}{c|}{Portfolio (c)} & \multicolumn{2}{c|}{Portfolio (d)} \\ \hline
Metric   & MSPM           & MSPM(w/o)      & MSPM           & MSPM(w/o) & MSPM           & MSPM(w/o)      & MSPM           & MSPM(w/o)      \\ \hline
DRR (\%) & \textbf{0.404} & 0.065          & \textbf{0.938} & 0.163     & \textbf{0.403} & 0.040          & \textbf{0.383} & 0.002          \\ \hline
ARR (\%) & \textbf{131.5} & 9.8            & \textbf{566.6} & 25.8      & \textbf{125.8} & 1.1            & \textbf{115.6} & -5.9           \\ \hline
MD (\%)  & -31.3          & \textbf{-31.0} & \textbf{-60.6} & -62.8     & -37.6          & \textbf{-36.7} & -37.6          & \textbf{-25.3} \\ \hline
SR       & \textbf{2.86}  & 0.61           & \textbf{4.18}  & 1.00      & \textbf{2.57}  & 0.32           & \textbf{2.45}  & 0.01           \\ \hline
\multicolumn{9}{|l|}{MSPM: EAM-enabled MSPM; MSPM(w/o): EAM-disabled MSPM}                                                                  \\ \hline
\end{tabular}
\label{tab:EAM_backtest}
\end{adjustwidth}
\end{table}

\subsubsection*{Discussion on scalability and reusability of MSPM}
To address the issue of inefficient model training in RL-based PM, EAMs are designed to be independent and reusable. Once an EAM has been trained, it can be added to any SAM without retraining. For example, in the previous sections, portfolio(a) and portfolio(b) share one EAM in common: GOOGL, and it saves time and resources from redundant model training.  On the other hand, to address the issues of ad-hoc and fixed model training in RL-based PM, MSPM allows the number of EAMs connected to any single SAM to be scaled up. In the EAM: Case study section, each EAM represents a single asset, and since these EAMs are trained, they are ready to be connected to any SAM. For example, to build a portfolio containing two assets, e.g., AAPL and TSLA, we can connect the corresponding two EAMs to an SAM to train and build the portfolio. Meanwhile, the rest of the EAMs can also be used in other portfolios. If later we want to scale up the volume of this portfolio to four assets, we simply add two more EAMs, e.g., GOOGL and NVDA, to the SAM without wasting time for training the EAMs again. Although SAM needs to be retrained once its volume is scaled up, the benefits brought by the EAMs are considerable since it has been validated in the previous section that the performance of an EAM-enabled SAM is largely improved compared to an EAM-disabled EAMs. Moreover, MSPM's scalability allows EAMs to accommodate the need for heterogeneous and alternative data input, like the sentiments data utilized in our research.
Therefore, with MSPM's scalability and reusability to create dynamic and adaptive portfolios, researchers and portfolio managers can simultaneously perform capital reallocation for various portfolios of a large volume of assets at scale by parallel computing.

\section*{Limitations and future work}
In this paper, to accommodate MSPM in sequential decision-making problems of PM, we only implement DQN and PPO to formalize the agents in EAM and SAM modules. We left the implementation of other algorithms in MSPM to future studies. Additionally, the trade-off between the stability of DRR and the performance metrics (ARR, DRR, or SR) may be further considered when designing the reward functions in future studies. We only implement the historical prices and sentiments data in this research, and we plan to utilize more heterogeneous data, e.g., satellite images, in the future studies.

\section*{Conclusion}
We propose MSPM, a modularized multi-agent RL-based system, to bring scalability and reusability to financial portfolio management. We design and develop two types of modules in MSPM: EAM and SAM. EAM is an asset-dedicated module that takes heterogeneous data and utilizes a DQN-based agent to generate signal-comprised information. On the other hand, SAM is a decision-making module that receives stacked information from the connected EAMs to reallocate the assets in a portfolio. As EAMs can be combined and connected to any SAMs at will, with this modularized and reusable design, MSPM addresses the issue of ad-hoc, fixed, and inefficient model training in the existing RL-based methods. By experimenting, we confirm that MSPM outperforms various baselines in terms of the accumulated rate of return, daily rate of return, and Sortino ratio. Additionally, to exemplify the high quality and reliability of the signals generated by EAM, we inspect the position-holding of five different EAMs. Furthermore, we validate the necessity of EAM by back-testing and comparing the EAM-enabled and disabled MSPMs on four different portfolios. The experimental results prove that MSPM is qualified as a stepping stone to inspire more creative system designs in reinforcement learning-based financial portfolio management.

\section*{Supporting information}

\paragraph*{S1 Appendix.}
\label{S1_Appendix}
{\bf Model selection for EAM and hyperparameter tuning.} For model selection, we have tested different architectures of neural network models for the DQN agent in EAM. Among them, we chose Residual Network with 1-D convolution since it performed the best on the validation dataset described in Table~\ref{tab:datasets} in Data ranges section. We also performed numerous experiments for hyperparameter tuning on the validation dataset to make sure the hyperparameters implemented and stated in the article are the optimized for the use cases in this research.

\paragraph*{\href{https://ndownloader.figstatic.com/files/34107811}{S2 Fig.}}
\label{S2_Fig}
{\bf 5-day RstdDRR of Portfolio(a): MSPM versus ARL.}

\paragraph*{\href{https://ndownloader.figstatic.com/files/34107814}{S3 Fig.}}
\label{S3_Fig}
{\bf 5-day RstdDRR of Portfolio(b): MSPM versus ARL.}

\paragraph*{\href{https://ndownloader.figstatic.com/files/34107817}{S4 Fig.}}
\label{S4_Fig}
{\bf Underwater plot of MSPM for Portfolio(a).}

\paragraph*{\href{https://ndownloader.figstatic.com/files/34107820}{S5 Fig.}}
\label{S5_Fig}
{\bf Underwater plot of ARL for Portfolio(a).}

\paragraph*{\href{https://ndownloader.figstatic.com/files/34107823}{S6 Fig.}}
\label{S6_Fig}
{\bf Underwater plot of MSPM for Portfolio(b).}

\paragraph*{\href{https://ndownloader.figstatic.com/files/34107826}{S7 Fig.}}
\label{S7_Fig}
{\bf Underwater plot of ARL for Portfolio(b).}


\begin{thebibliography}{10}

\bibitem{10.2307/2975974}
Markowitz H.
\newblock Portfolio Selection.
\newblock The Journal of Finance. 1952;7(1):77--91.

\bibitem{jiang2017deep}
Jiang Z, Xu D, Liang J. A Deep Reinforcement Learning Framework for the
  Financial Portfolio Management Problem; 2017.
\newblock Available from: \url{https://arxiv.org/abs/1706.10059}.

\bibitem{10.5555/3044805.3044850}
Silver D, Lever G, Heess N, Degris T, Wierstra D, Riedmiller M.
\newblock Deterministic Policy Gradient Algorithms.
\newblock In: Proceedings of the 31st International Conference on International
  Conference on Machine Learning - Volume 32. ICML'14. JMLR.org; 2014. p.
  I–387–I–395.

\bibitem{lillicrap2019continuous}
Lillicrap TP, Hunt JJ, Pritzel A, Heess N, Erez T, Tassa Y, et~al.. Continuous
  control with deep reinforcement learning; 2019.
\newblock Available from: \url{https://arxiv.org/abs/1509.02971}.

\bibitem{liang2018adversarial}
Liang Z, Chen H, Zhu J, Jiang K, Li Y. Adversarial Deep Reinforcement Learning
  in Portfolio Management; 2018.
\newblock Available from: \url{https://arxiv.org/abs/1808.09940}.

\bibitem{schulman2017proximal}
Schulman J, Wolski F, Dhariwal P, Radford A, Klimov O. Proximal Policy
  Optimization Algorithms; 2017.
\newblock Available from: \url{https://arxiv.org/abs/1707.06347}.

\bibitem{Ye_Pei_Wang_Chen_Zhu_Xiao_Li_2020}
Ye Y, Pei H, Wang B, Chen PY, Zhu Y, Xiao J, et~al.
\newblock Reinforcement-Learning Based Portfolio Management with Augmented
  Asset Movement Prediction States.
\newblock Proceedings of the AAAI Conference on Artificial Intelligence.
  2020;34(01):1112--1119.
\newblock doi:{10.1609/aaai.v34i01.5462}.

\bibitem{10.1007/978-3-642-12433-4_7}
L{\'o}pez VF, Alonso N, Alonso L, Moreno MN.
\newblock A Multiagent System for Efficient Portfolio Management.
\newblock In: Demazeau Y, Dignum F, Corchado JM, Bajo J, Corchuelo R, Corchado
  E, et~al., editors. Trends in Practical Applications of Agents and Multiagent
  Systems. Berlin, Heidelberg: Springer Berlin Heidelberg; 2010. p. 53--60.

\bibitem{Sycara-1995-14017}
Sycara K, Decker K, Zeng D.
\newblock Designing a Multi-Agent Portfolio Management System.
\newblock In: Proceedings of the AAAI Workshop on Internet Information Systems;
  1995.

\bibitem{Lee_2020}
Lee J, Kim R, Yi SW, Kang J.
\newblock MAPS: Multi-Agent reinforcement learning-based Portfolio management
  System.
\newblock Proceedings of the Twenty-Ninth International Joint Conference on
  Artificial Intelligence. 2020;doi:{10.24963/ijcai.2020/623}.

\bibitem{mnih2013atari}
Mnih V, Kavukcuoglu K, Silver D, Graves A, Antonoglou I, Wierstra D, et~al..
  Playing Atari with Deep Reinforcement Learning; 2013.
\newblock Available from: \url{https://arxiv.org/abs/1312.5602}.

\bibitem{ensemble}
Rokach L.
\newblock Ensemble-based classifiers.
\newblock Artif Intell Rev. 2010;33:1--39.
\newblock doi:{10.1007/s10462-009-9124-7}.

\bibitem{Liu_Liu_Zhao_Pan_Liu_2020}
Liu Y, Liu Q, Zhao H, Pan Z, Liu C.
\newblock Adaptive Quantitative Trading: An Imitative Deep Reinforcement
  Learning Approach.
\newblock Proceedings of the AAAI Conference on Artificial Intelligence.
  2020;34(02):2128--2135.
\newblock doi:{10.1609/aaai.v34i02.5587}.

\bibitem{10.1093/rfs/hhm075}
DeMiguel V, Garlappi L, Uppal R.
\newblock {Optimal Versus Naive Diversification: How Inefficient is the 1/N
  Portfolio Strategy?}
\newblock The Review of Financial Studies. 2007;22(5):1915--1953.
\newblock doi:{10.1093/rfs/hhm075}.

\bibitem{Fama1970EFFICIENTCM}
Fama EF.
\newblock EFFICIENT CAPITAL MARKETS: A REVIEW OF THEORY AND EMPIRICAL WORK*.
\newblock Journal of Finance. 1970;25:383--417.

\bibitem{TechnicalAnalysis}
Murphy JJ.
\newblock Technical Analysis of the Financial Markets: A Comprehensive Guide to
  Trading Methods and Applications.
\newblock New York Institute of Finance; 1999.

\bibitem{QuoteMedia}
QuoteMedia. End-Of-Day Data; 2020.
\newblock \url{https://data.nasdaq.com/data/EOD-end-of-day-us-stock-prices}.

\bibitem{NS1}
InfoTrie. FinSentS Web News Sentiment; 2021.
\newblock \url{https://data.nasdaq.com/databases/NS1/data}.

\bibitem{araci2019finbert}
Araci D. FinBERT: Financial Sentiment Analysis with Pre-trained Language
  Models; 2019.
\newblock Available from: \url{https://arxiv.org/abs/1908.10063}.

\bibitem{devlin-etal-2019-bert}
Devlin J, Chang MW, Lee K, Toutanova K.
\newblock {BERT}: Pre-training of Deep Bidirectional Transformers for Language
  Understanding.
\newblock In: Proceedings of the 2019 Conference of the North {A}merican
  Chapter of the Association for Computational Linguistics: Human Language
  Technologies, Volume 1 (Long and Short Papers). Minneapolis, Minnesota:
  Association for Computational Linguistics; 2019. p. 4171--4186.

\bibitem{10.5555/3279266}
Lapan M.
\newblock Deep Reinforcement Learning Hands-On: Apply Modern RL Methods, with
  Deep Q-Networks, Value Iteration, Policy Gradients, TRPO, AlphaGo Zero and
  More.
\newblock Packt Publishing; 2018.

\bibitem{8331585}
{Hong} S, {Wu} M, {Zhou} Y, {Wang} Q, {Shang} J, {Li} H, et~al.
\newblock ENCASE: An ENsemble ClASsifiEr for ECG classification using expert
  features and deep neural networks.
\newblock In: 2017 Computing in Cardiology (CinC); 2017. p. 1--4.

\bibitem{wang2016dueling}
Wang Z, Schaul T, Hessel M, van Hasselt H, Lanctot M, de~Freitas N. Dueling
  Network Architectures for Deep Reinforcement Learning; 2016.
\newblock Available from: \url{https://arxiv.org/abs/1511.06581}.

\bibitem{10.5555/3016100.3016191}
Hasselt Hv, Guez A, Silver D.
\newblock Deep Reinforcement Learning with Double Q-Learning.
\newblock In: Proceedings of the Thirtieth AAAI Conference on Artificial
  Intelligence. AAAI'16. AAAI Press; 2016. p. 2094–2100.

\bibitem{10.5555/3104322.3104425}
Nair V, Hinton GE.
\newblock Rectified Linear Units Improve Restricted Boltzmann Machines.
\newblock In: Proceedings of the 27th International Conference on International
  Conference on Machine Learning. ICML'10. Madison, WI, USA: Omnipress; 2010.
  p. 807–814.

\bibitem{doi:10.1080/14697688.2011.570368}
Ormos M, Urbán A.
\newblock Performance analysis of log-optimal portfolio strategies with
  transaction costs.
\newblock Quantitative Finance. 2013;13(10):1587--1597.
\newblock doi:{10.1080/14697688.2011.570368}.

\bibitem{Sortino59}
Sortino FA, Price LN.
\newblock Performance Measurement in a Downside Risk Framework.
\newblock The Journal of Investing. 1994;3(3):59--64.
\newblock doi:{10.3905/joi.3.3.59}.

\bibitem{10.1145/2512962}
Li B, Hoi SCH.
\newblock Online Portfolio Selection: A Survey.
\newblock ACM Comput Surv. 2014;46(3).
\newblock doi:{10.1145/2512962}.

\bibitem{mlfinlab}
{Hudson and Thames Quantitative Research}. Machine Learning Financial
  Laboratory (MlFinLab); 2021.
\newblock \url{https://github.com/hudson-and-thames/mlfinlab}.

\bibitem{ADictionaryofFinanceandBanking}
Law J, Smullen J.
\newblock A Dictionary of Finance and Banking.
\newblock Oxford University Press; 2008.
\newblock Available from:
  \url{https://www.oxfordreference.com/view/10.1093/acref/9780199229741.001.0001/acref-9780199229741}.

\bibitem{10.1093/biomet/52.3-4.591}
SHAPIRO SS, WILK MB.
\newblock {An analysis of variance test for normality (complete samples)†}.
\newblock Biometrika. 1965;52(3-4):591--611.
\newblock doi:{10.1093/biomet/52.3-4.591}.

\bibitem{levene}
Olkin I, Hotelling H.
\newblock Contributions to probability and statistics: Essays in honor of
  Harold Hotelling.
\newblock Stanford University Press; 1960.

\bibitem{10.2307/2236101}
Mann HB, Whitney DR.
\newblock On a Test of Whether one of Two Random Variables is Stochastically
  Larger than the Other.
\newblock The Annals of Mathematical Statistics. 1947;18(1):50--60.

\end{thebibliography}

\end{document}